\documentclass{aa}
\usepackage{graphicx}
\begin{document}

\title{Photometric and kinematic studies of open star clusters}
\subtitle{III. NGC\,4103, NGC\,5281, and NGC\,4755\thanks{Partly based on
    observations from ESO, La Silla} \fnmsep \thanks{Tables
    \ref{n4103cdsphot}, \ref{n5281cdsphot}, \ref{n4755cdsphot},
    \ref{n4103cdseb}, \ref{n5281cdseb}, and \ref{n4755cdseb} are only
    available in electronic form at the CDS via anonymous ftp to
    cdsarc.u-strasbg.fr (130.79.125.5) or via
    http://cdsweb.u-strasbg.fr/cgi-bin/qcat?J/A+A/}}

\author{J\"org Sanner\inst{1} \and Jens Brunzendorf\inst{2} \and
  Jean-Marie Will\inst{1} \fnmsep \thanks{\emph{Present address:}
  Hewlett-Packard GmbH, Schickardstr. 25, 71034 B\"oblingen, Germany} \and
  Michael Geffert\inst{1}}

\institute{Sternwarte der Universit\"at Bonn, Auf dem H\"ugel 71, 
           53121 Bonn, Germany \and
           Th\"uringer Landessternwarte Tautenburg, Sternwarte 5,
           07778 Tautenburg, Germany}

\offprints{J\"org Sanner, \email{jsanner@astro.uni-bonn.de}}

\date{Received 4 November 2000 / Accepted 29 January 2001}

\titlerunning{Photometric and kinematic studies of open star clusters. III}
\authorrunning{J. Sanner et~al.}

\abstract{We present CCD photometry and proper motion studies of the three
  open star clusters NGC\,4103, NGC\,5281, and NGC\,4755 ($\kappa$ Cru). By
  fitting isochrones to the colour magnitude diagrams, we found that all three
  objects are young open star clusters with ages of at most $t=45 \mbox{~Myr}$. They are located at distances from approx. $1600 \mbox{~pc}$ to
  $2200 \mbox{~pc}$, derived from distance moduli $(m-M)_0$ ranging from $11
  \mbox{~mag}$ to $12 \mbox{~mag}$. We combined membership determinations
  based on proper motions and statistical field star subtraction to derive the
  initial mass function (IMF) of the clusters. The shape of the IMFs could be
  represented by power laws with exponents of $\Gamma=-1.46 \pm 0.22$ for
  NGC\,4103, $\Gamma=-1.60 \pm 0.50$ for NGC\,5281, and $\Gamma=-1.68 \pm
  0.14$ for NGC\,4755, when -- as a reference -- Salpeter's (\cite{salpeter})
  value would be $\Gamma=-1.35$. These results agree well with other IMF
  studies of open star clusters.
\keywords{open clusters and associations: individual: NGC\,4103; NGC\,5281;
          NGC\,4755 -- astrometry -- stars: kinematics -- Hertzsprung-Russell
          and C-M diagrams -- stars: luminosity function, mass function}}

\maketitle

\section{Introduction}

We continue our series of studies of open star clusters and present results on
three clusters of the southern hemisphere, NGC\,4103, NGC\,5281, and NGC\,4755
($\kappa$ Cru).

Our goals are to determine the basic parameters age, distance, and reddening
of the clusters, and the luminosity functions (LFs) and initial mass functions
(IMFs) of the target objects. For reviews of the IMF see, e.g., Scalo
(\cite{scalo1,scalo2}). Restricted to certain mass intervals, the IMF can be
described by a power law in the form $\mbox{d} \log N(m) \sim m^\Gamma
\mbox{d} \log m$. In this notation the ``classical'' value found by Salpeter
(\cite{salpeter}) for the Solar neighbourhood would be
$\Gamma=-1.35$. According to Scalo (\cite{scalo2}), we can expect our values
for $\Gamma$ to be located within the interval $[1.2;2.2]$, a range which is
valid for the intermediate mass range of $1 M_\odot \la m \la 10 M_\odot$.

For open star clusters the number of field stars in the region of the cluster
is normally of the same order of magnitude as the number of cluster
members. Therefore it is essential to subtract the field stars before
determining the IMF to obtain a reliable mass function. This subtraction is
performed by a combination of membership determination with stellar proper
motions and a statistical field star subtraction on the basis of the colour
magnitude diagrams (CMDs).

NGC\,4103 is located at $\alpha_{2000}=12^{\rm h} 6^{\rm m} 43^{\rm s}$,
$\delta_{2000}=-61^{\circ} 15 \arcmin$. It is dominated by a handful of
moderately bright stars with magnitudes around $V \approx 10 - 12 \mbox{~mag}$ and is not very richly populated. According to the Lyng{\aa}
(\cite{lynga}) catalogue, it has a diameter of $5 \arcmin$ and an age of $7.5$
Myr. Besides some earlier publications on photographic photometry (Becker
\cite{becker}, Wesselink \cite{wesselink}), the cluster was recently studied
with CCD photometry by Sagar \& Cannon (\cite{sagar4103}). Proper motions were
studied before by King (\cite{king4103}). With our work, we will be able to
improve the previous results: We reach fainter limiting magnitudes than the
previous photometric studies. For the astrometric part, we have a larger
difference between first and second epoch plate material. In addition, we had
the possibility to digitise and reduce entire photographic plates, which
results in more precise and reproducible data. We therefore expect a higher
precision of our results than in the earlier studies. Finally, the combination
of photometry and astrometry will provide a better understanding of the stars
which physically belong to the cluster. Our work is the first IMF study for
NGC\,4103.

NGC\,5281 is located at $\alpha_{2000}=13^{\rm h} 46^{\rm m} 37^{\rm s}$,
$\delta_{2000}=-62^\circ 54\arcmin$. Four bright (up to
$V \simeq 6.6 \mbox{~mag}$) stars almost located in one line on the
sky are striking. No large population of faint stars is visible. One only
finds sparse photometric data on this cluster, e.g.\ the photoelectric
photometry of 18 stars by Moffat \& Vogt (\cite{moffatvogt}). It has not yet
been the target of CCD photometry or a proper motion study.

NGC\,4755 (also known as $\kappa$ Cru or Herschel's Jewel Box) is located at
$\alpha_{2000}=12^{\rm h} 53^{\rm m} 49^{\rm s}$, $\delta_{2000}=-60^\circ 23
\arcmin$. The cluster is mentioned with an angular diameter of $10 \arcmin$ in
Lyng{\aa} (\cite{lynga}). It has been the subject of various studies before,
including CCD photometry (Sagar \& Cannon \cite{sagar4755}) and proper motion
studies (King \cite{king4755}). However, the same reasons as for
NGC\,4103 hold for revisiting this object. The IMF of NGC\,4755 has not yet
been investigated. With approx.\ 10 stars with $V<10 \mbox{~mag}$, NGC\,4755
is an easy and prominent object in the southern hemisphere even for
binoculars. The designation ``Jewel Box'' results from the bright stars with
one of them being very red ($B-V \approx 2.2 \mbox{~mag}$) whereas the others
show blue colours with $B-V \la 0.3 \mbox{~mag}$.

In Sects.~\ref{suedphoto} and \ref{suedastro}, we present the data and
techniques used for our photometric and kinematic studies and describe the
basic results of our work, followed by the analysis of the individual
clusters, including the IMF determinations, in
Sect.~\ref{suedcmds}. Concluding remarks are given in Sect.~\ref{suedsumm}.

\section{Photometry} \label{suedphoto}

The CCD observations of NGC\,4103 and NGC\,5281 were made from April 25 to 30,
1994, with the 1.54~m Danish telescope at ESO, La Silla, equipped with DFOSC
and the $1 \mbox{~K} \times 1 \mbox{~K}$ pixel CCD chip ESO \# 28. The pixel
size of the detector is $24 \mbox{~} \mu \mbox{m}$. The resolution of this
configuration is $0 \farcs 375 \mbox{~pix}^{-1}$ resulting in a field of view
of $6 \farcm 4 \times 6 \farcm 4$. The observations were done using Bessell
$B$, $V$, and $I$ filters. The seeing during all observations was of the order
of $1 \farcs 0$ to $1 \farcs 2$. The NGC\,4755 data were taken with the same
telescope only in $B$ and $V$ filters from February 5 to 8, 1996,
with a LORAL $2 \mbox{~K} \times 2 \mbox{~K}$ chip with a pixel
size of $15 \mbox{~} \mu \mbox{m}$. Since the outer regions of the images
showed only a very limited signal, the usable field of view is approx.\ $1800
\times 1800 \mbox{~pix}$ or $9 \arcmin \times 9 \arcmin$. The seeing of the
observations differed from $1 \farcs 3$ to almost $2 \arcsec$, a value which
was reached in one night only.

To cover a sufficiently large area for the cluster and additional field
star region, two (NGC\,5281, NGC\,4755) or three (NGC\,4103) CCD fields were
observed. These fields were observed in consecutive nights. At least one of
the observations of each cluster could be proved to have taken place in
photometric conditions. Tables \ref{suedaufnahmen} and \ref{n4755aufnahmen}
give an overview of the observations used for this work. 

\begin{table}
\caption[]{\label{suedaufnahmen} Overview of the images used for the CCD
  photometry of NGC\,4103 and NGC\,5281. Multiple observations of the same exposure time were added before data reduction}
\begin{tabular}{lcc@{\hspace{1cm}}rrr}
\hline
cluster & field & filter & \multicolumn{3}{c}{$t_{\rm exp} [{\rm s}]$} \\
\hline
NGC 4103 & A & $B$ & $2$ & $5 \times 180$ & $2 \times 900$\\
         &   & $V$ & $0.5$ & $5 \times 30$ & $5 \times 300$ \\
         &   & $I$ & $0.3$ & $3$ & $4 \times 180$\\
         & B & $B$ & $1$ & $60$ & $2 \times 1200$ \\
         &   & $V$ & $1$ & $60$ & $2 \times 600$ \\
         &   & $I$ & $0.5$ & $10$ & $2 \times 600$ \\
         & C & $B$ & $1$ & $5$ & $2 \times 1200$ \\
         &   & $V$ & $1$ & $60$ & $2 \times 600$ \\
         &   & $I$ & $1$ & $2$ & $2 \times 600$ \\
\hline
NGC 5281 & A & $B$ & $0.1$ & $ 30$ & $4 \times 600$ \\
         &   & $V$ & $0.1$ & $ 30$ & $4 \times 300$ \\
         &   & $I$ & $0.1$ & $ 20$ & $4 \times 300$ \\
         & B & $B$ & $  2$ & $120$ & $2 \times 1200$ \\
         &   & $V$ & $  1$ & $ 60$ & $2 \times 600$ \\
         &   & $I$ & $  1$ & $ 60$ & $2 \times 600$ \\
\hline
\end{tabular}
\end{table}

\begin{table}
\caption[]{\label{n4755aufnahmen} Overview of the images used for the CCD
  photometry of NGC\,4755. Multiple observations of the same exposure time
  were added before data reduction}
\begin{tabular}{cc@{\hspace{1cm}}rrrrr}
\hline
field & filter & \multicolumn{5}{c}{$t_{\rm exp} [{\rm s}]$}\\
\hline
A & $B$ & $3$ & $10$ & $15$ & $180$ & $1800$ \\
  & $V$ & $1$ & $ 3$ &  $2 \times 5$ & $2 \times 60$ & $600$\\
B & $B$ & $3$ & $10$ & $15$ & $180$ & $1800$ \\
  & $V$ &     & $ 2$ & $2 \times 5$ & $60$ & $600$ \\
\hline
\end{tabular}
\end{table}

After standard image processing and addition of the images with equal exposure
times, PSF photometry was performed with DAOPHOT II (Stetson \cite{daophot})
running under IRAF. Before combining the results of the images of different
exposure times we deselected objects with too high errors, sharpness, and
$\chi$ values according to the DAOPHOT output. The data of one field per
cluster were calibrated from instrumental to Johnson magnitudes using standard
fields from Landolt (\cite{landolt}) which were observed in addition to the
objects. The other field(s) were calibrated with the help of the stars in the
overlapping regions of the different fields.

After correction of the exposure time to $1 \mbox{~s}$ we determined the
parameters necessary for the calibration from the following equations:
\begin{eqnarray}
v &=& V + z_V + k_V \cdot X_V + c_V \cdot (B-V) \label{eichv}\\
b &=& B + z_B + k_B \cdot X_B + c_B \cdot (B-V) \label{eichb}\\
i &=& I + z_I + k_I \cdot X_I + c_I \cdot (V-I), \label{eichi}
\end{eqnarray}
where $X$ stands for the airmass, capital letters for apparent and lower case
letters for instrumental magnitudes. The values for the zero points $z$
and colour coefficients $c$ are listed in Table \ref{suedkoeff}, the
extinction coefficients $k$ were provided by the Geneva Observatory Photometric
Group (see Burki et~al.\ \cite{extinkt}). In all nights the extinction was of
the order of $k_B=0.25$, $k_V=0.14$, and $k_I=0.06$, respectively.

\begin{table}
\caption[]{\label{suedkoeff} Parameters used for the calibration of the
  photometric data of NGC\,4103, NGC\,5281, and NGC\,4755. The parameters
  are defined by Eqs.\ (\ref{eichv}) to (\ref{eichi}). For
  NGC\,4755 only $B$ and $V$ data were observed}
\begin{tabular}{lrrr}
\hline
 & \multicolumn{1}{c}{NGC\,4103} & \multicolumn{1}{c}{NGC\,5281} & \multicolumn{1}{c}{NGC\,4755}\\
 & \multicolumn{1}{c}{April 26, 1994} & \multicolumn{1}{c}{April 28, 1994} &
 \multicolumn{1}{c}{February 5, 1996}\\\hline
$z_V$  & $ 1.689 \pm 0.045$ & $ 1.676 \pm 0.014$ & $ 1.331 \pm 0.097$ \\
$c_V$  & $-0.091 \pm 0.036$ & $-0.076 \pm 0.015$ & $-0.082 \pm 0.019$ \\
$z_B$  & $ 2.232 \pm 0.015$ & $ 2.184 \pm 0.006$ & $ 2.082 \pm 0.120$ \\
$c_B$  & $-0.188 \pm 0.012$ & $-0.162 \pm 0.006$ & $-0.231 \pm 0.018$ \\
$z_I$  & $ 2.592 \pm 0.024$ & $ 2.598 \pm 0.012$ & \\
$c_I$  & $ 0.020 \pm 0.016$ & $-0.012 \pm 0.011$ & \\
\hline
\end{tabular}
\end{table}

\subsection{NGC\,4103} \label{n4103photo}

\begin{figure}
\centerline{
\includegraphics[width=\hsize]{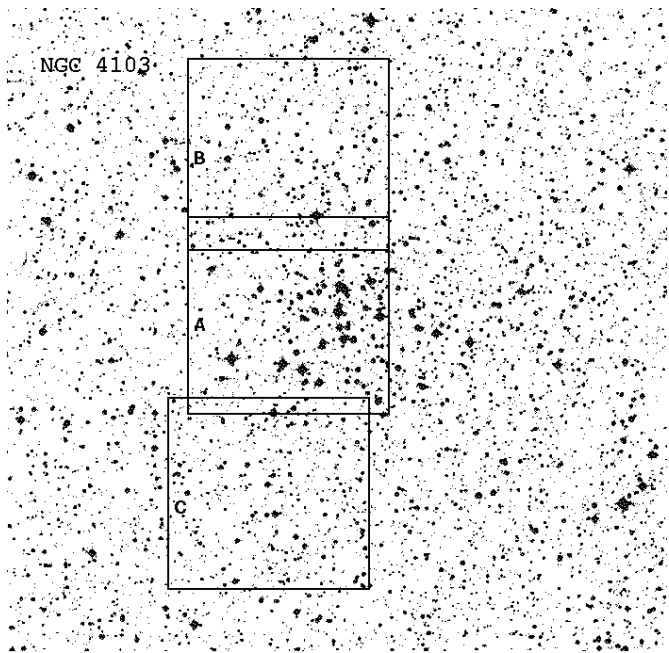}
}
\caption[]{\label{n4103bild} Region of NGC\,4103 taken from the
  DSS. The field of view is $20\arcmin \times 20 \arcmin$. North is up and
  east to the left. The three CCD fields A, B, and C observed for this
  work are indicated, measuring $6 \farcm 4 \times 6 \farcm 4$}
\end{figure}

Three fields around this cluster were observed: the first one -- hereafter
referred to as ``field A'' -- centred slightly east of the cluster's centre,
the second and third ones -- ``fields B and C'' -- pointing $5 \arcmin$
north and south of this position, respectively. Fields B and C were observed
to provide a sufficiently large area which does not contain cluster
stars, which is necessary for the statistical field star
subtraction. Figure \ref{n4103bild} shows a Digitized Sky Survey (DSS) image of
the region of NGC\,4103 with the location of the three fields marked.

Transformation of instrumental to apparent magnitudes was performed for field
C, which was observed under photometric conditions.

Typical internal errors of stars in different magnitude ranges are given in
Table \ref{n4103fehler}. In addition, we compared our data with the results of
two previous studies: A comparison with the data of Wesselink
(\cite{wesselink}) showed standard deviations of the order of $0.03
\mbox{~mag}$ in all $V$, $B-V$, and $V-I$; the corresponding values for the
brighter ($V<16 \mbox{~mag}$) stars of the study of Sagar \& Cannon
(\cite{sagar4103}) are $0.04 \mbox{~mag}$ in $V$, $0.06 \mbox{~mag}$ in $B-V$,
and $0.02 \mbox{~mag}$ in $V-I$. The entire photometric data are listed in
Table \ref{n4103cdsphot}. Besides our own numbering system, we added the
  stellar numbers of Wesselink (\cite{wesselink}) as a reference.

\begin{table}
\caption[]{\label{n4103fehler} Typical photometric errors of stars in
  NGC\,4103 in different magnitude ranges. The values given are derived from
  the DAOPHOT II output for the accuracy of the PSF photometry}
\begin{tabular}{lccc}
\hline
 & $\Delta V$ & $\Delta (B-V)$ & $\Delta (V-I)$ \\
 & [mag] & [mag] & [mag]\\
\hline
$V<15 \mbox{~mag}$ & $0.02$ & $0.03$ & $0.04$ \\
$V>15 \mbox{~mag}$ & $0.04$ & $0.06$ & $0.05$ \\
\hline
\end{tabular}
\end{table}

\refstepcounter{table} \label{n4103cdsphot}

\begin{figure*}
\centerline{
\includegraphics[width=\textwidth]{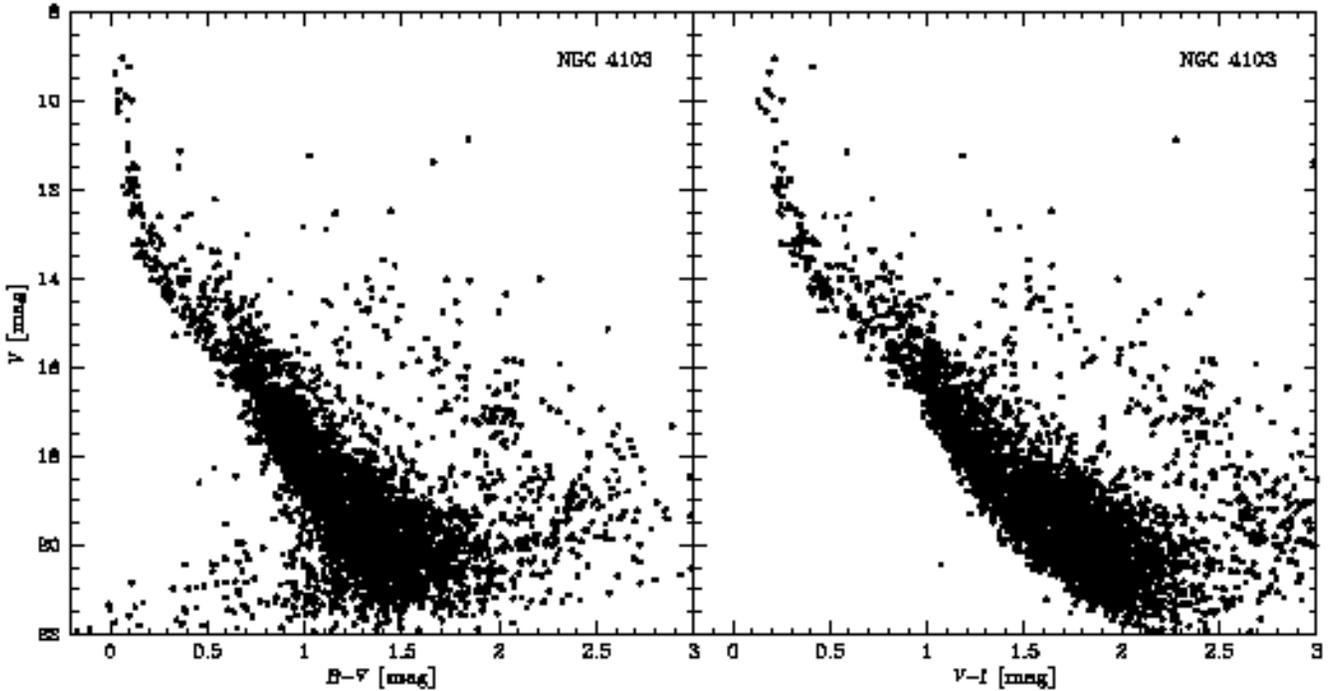}
}
\caption[]{\label{n4103cmds} Colour magnitude diagrams of all stars detected
  in fields A, B, and C of NGC\,4103. Note that these diagrams are still
  contaminated with field stars; for cleaned CMDs see
  Fig.~\ref{n4103cmdshaufen}}
\end{figure*}

From all data of the three fields, we derived the $(B,V)$ and $(V,I)$ colour
magnitude diagrams which we present in Fig.~\ref{n4103cmds}. The lack of bright
red stars already shows at this point that NGC\,4103 is a young star
cluster. Around $V \approx 11-15$ mag, an indication for a secondary main
sequence is visible. From $V=14 \mbox{~mag}$ downwards, another scattered
feature approximately parallel to the main sequence, but $1 \mbox{~mag}$
redder, is visible. These stars likely belong to the field population. As the
CMDs are still contaminated with field stars, we postpone the analysis of the
properties of the cluster to Sect.~\ref{n4103sect}.

\subsection{NGC\,5281} \label{n5281photometry}

\begin{figure}
\centerline{
\includegraphics[width=\hsize]{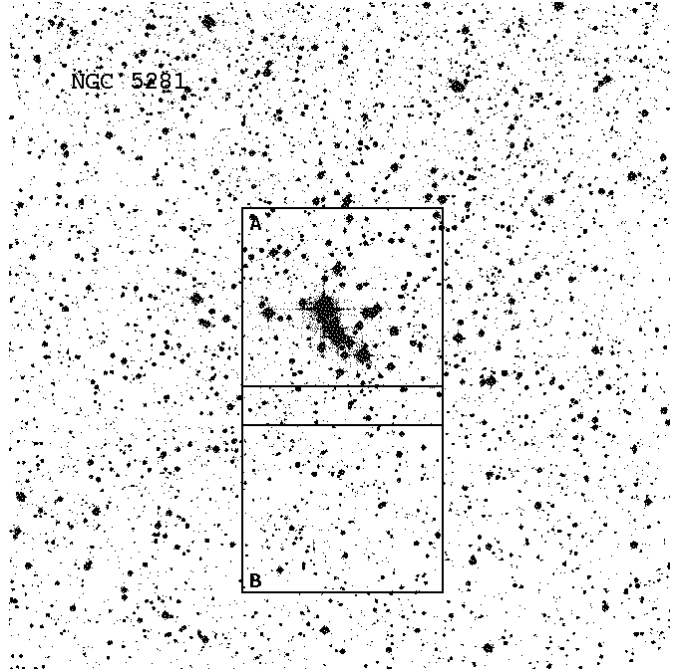}
}
\caption[]{\label{n5281bild} NGC\,5281. The characteristics of this DSS image
  are the same as for Fig.~\ref{n4103bild}. The two CCD fields A and B are
  indicated}
\end{figure}

With an angular diameter of $7\arcmin$ according to the Lyng{\aa}
(\cite{lynga}) catalogue, the object is slightly larger than the field of
view of the telescope/detector configuration used for the photometric
observations. Besides a pointing with the cluster in its centre (``field
A''), we observed a region located $6\arcmin$ south of the cluster
(``field B''), see Fig.~\ref{n5281bild}.

\begin{figure*}
\centerline{
\includegraphics[width=\textwidth]{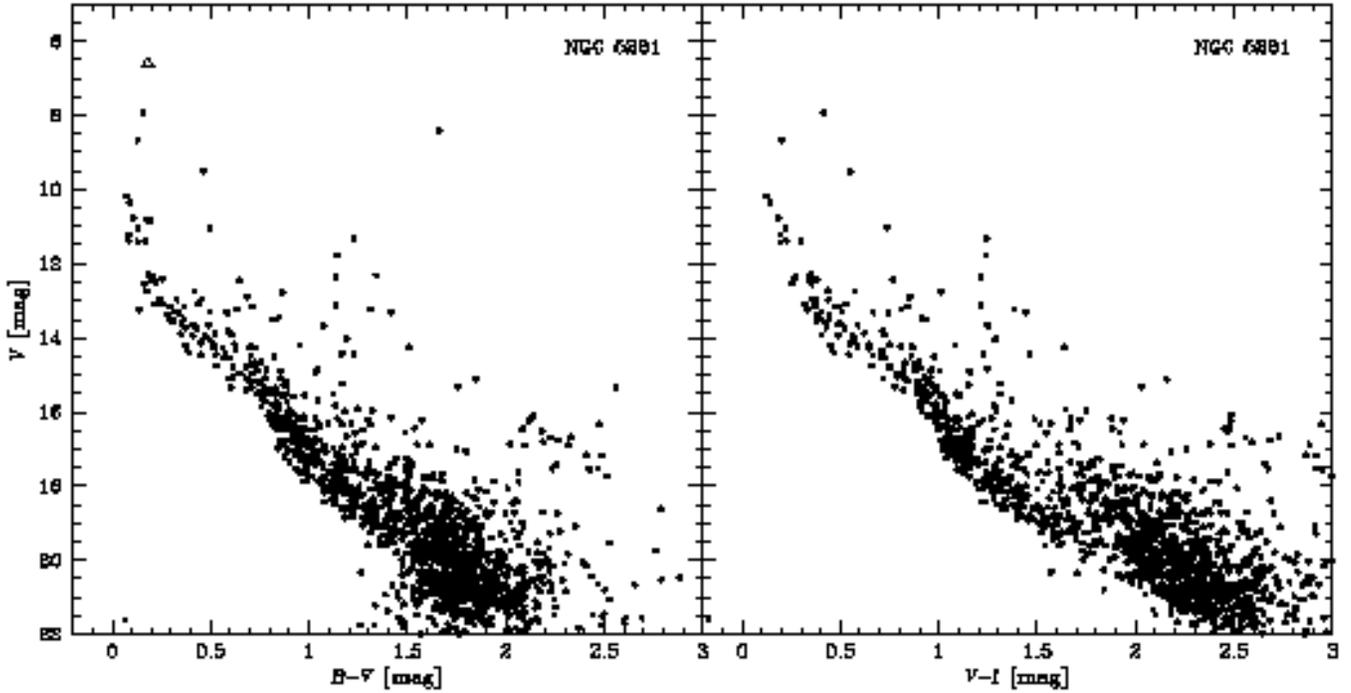}
}
\caption[]{\label{n5281cmds} Colour magnitude diagrams of all stars in the
  two fields of NGC\,5281. The brightest star in the $(B,V)$ CMD, marked with
  a triangle, was saturated even on the shortest exposures. Its magnitude and
  colour was adopted from Moffat \& Vogt (\cite{moffatvogt}). No $I$
  information is available for this object. These diagrams include field
  stars, cleaned diagrams are shown in Fig.~\ref{n5281cmdshaufen}}
\end{figure*}

The first night, when field A was observed, was excellent, but the second one
was not photometric. The values for the parameters $z$ and $c$ from
Eqs.\ (\ref{eichv}) to (\ref{eichi}) are listed in Table \ref{suedkoeff}. The
typical photometric errors (see Table \ref{n5281fehler}) show an uncommon
behaviour: Down to $V \approx 13 \mbox{~mag}$, the errors rise up to a few
hundredth of a magnitude to drop again in the magnitude range of $V \approx
13$ to $18 \mbox{~mag}$. A possible explanation is that the brightest
stars could only be measured on the shortest exposures on which the
photometric errors increase for brighter stars than on the longer
exposures. We could have applied a more stringent error selection, however we
accepted these uncertainties to avoid incompleteness in the
corresponding magnitude range. An external check of the results with the
photoelectric sequence of Moffat \& Vogt (\cite{moffatvogt}) showed
coincidence within $\Delta V=0.03 \mbox{~mag}$ and $\Delta (B-V)=0.06 \mbox{~mag}$ for 15 stars. The brightest star in the field ($V=6.61 \mbox{~mag}$,
$B-V=0.18 \mbox{~mag}$ according to Moffat \& Vogt (\cite{moffatvogt}) could
not be measured since it was overexposed even on the shortest images. We
manually added this star in the CMDs presented in Fig.~\ref{n5281cmds} and
marked it with a triangle. As these diagrams are still contaminated with field
stars, we will proceed with the analysis in Sect.~\ref{n5281sect}. A list of
the entire photometry is presented in Table \ref{n5281cdsphot} with the
star numbers according to Moffat \& Vogt (\cite{moffatvogt}) included.

\begin{table}
\caption[]{\label{n5281fehler} Typical photometric errors of stars in
  NGC\,5281 in different magnitude ranges. The values given are derived from
  the DAOPHOT II output for the accuracy of the PSF photometry. For a
  discussion of the comparably high errors for the bright stars, see
  Sect.~\ref{n5281photometry}}
\begin{tabular}{lccc}
\hline
 & $\Delta V$ & $\Delta (B-V)$ & $\Delta (V-I)$ \\
 & [mag] & [mag] & [mag]\\
\hline
$V<13 \mbox{~mag}$ & $0.04$ & $0.06$ & $0.05$ \\
$13<V<18 \mbox{~mag}$ & $0.02$ & $0.03$ & $0.03$ \\
$V>18 \mbox{~mag}$ & $0.06$ & $0.12$ & $0.07$ \\
\hline
\end{tabular}
\end{table}

\refstepcounter{table} \label{n5281cdsphot}

\subsection{NGC\,4755} \label{n4755photo}

\begin{figure}
\centerline{
\includegraphics[width=\hsize]{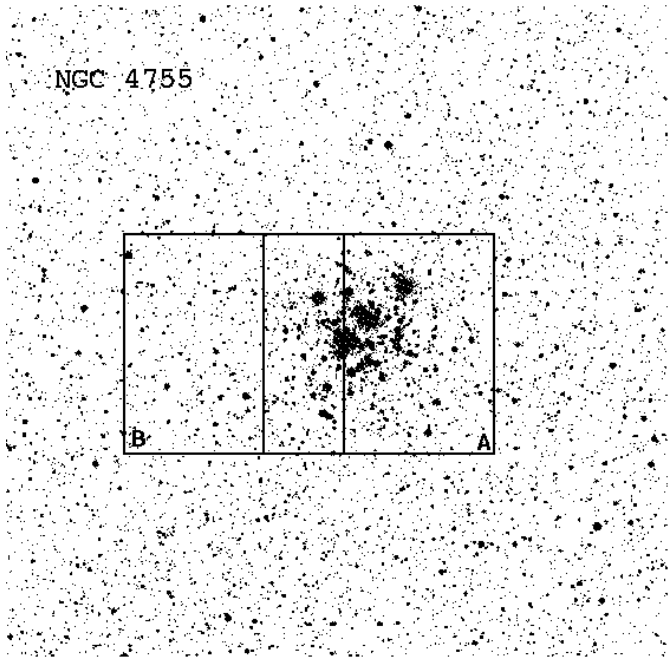}
}
\caption[]{\label{n4755bild} NGC\,4755. This excerpt from DSS shows a field of
  view of $30 \arcmin \times 30 \arcmin$ with north up and east to the
  left. The two CCD fields of our study are marked. Their field of view is
  approx.\ $9 \arcmin$}
\end{figure}

The diameter of NGC\,4755 is given as $10 \arcmin$ (Lyng{\aa}
\cite{lynga}). Nevertheless, it was sufficient to observe two fields as we
could prove with star counts. ``Field A'' has the cluster in its centre, and
``field B'' is located $8 \arcmin$ east of this position, see
Fig.~\ref{n4755bild}.

After calibrating field A with Landolt (\cite{landolt}) standard stars, we
found that with a linear and rotational transformation of the $(x,y)$
coordinates the stars in the region overlapping with field B could not well be
cross-identified: In case the tolerances were chosen too restrictively,
identification only worked out in a small region of the entire overlap,
whereas increasing the tolerances led to mis-identification of numerous
stars. This phenomenon is most likely caused by distortion in the outer
regions of the CCD field. Therefore we decided to adjust the magnitudes and
coordinates of field B to field A only with the help of the brightest stars:
With their low density the tolerances could be increased without the problem
of wrong identification. For our further analyses we used the entire field A
and the part of field B which does not overlap with field A to avoid stars to
be taken into account twice or not at all.

\begin{figure}
\centerline{
\includegraphics[width=\hsize]{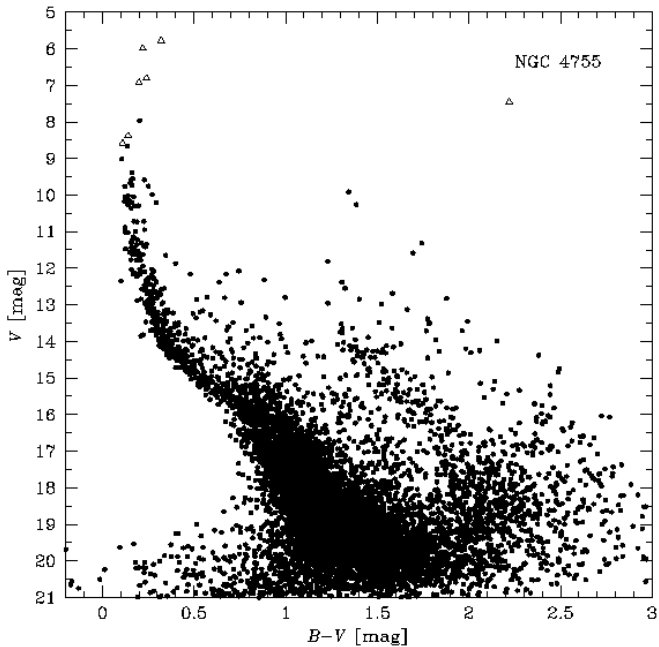}
}
\caption[]{\label{n4755cmd} Colour magnitude diagram of all stars of the two
  fields observed for NGC\,4755. The stars marked with triangles were
  overexposed even in our shortest exposures; $V$ and $B-V$ of these objects
  were excerpted from Dachs \& Kaiser (\cite{dachs})}
\end{figure}

The average internal photometric error is of the order of $ \la 0.01 \mbox{~mag}$ for stars brighter than $V \approx 17 \mbox{~mag}$ in both $V$ and
$B-V$  and increases below this point towards values of $0.03 \mbox{~mag}$ in
$V$ and $0.08 \mbox{~mag}$ in $B-V$, respectively. The photoelectric
photometry of Dachs \& Kaiser (\cite{dachs}) agrees very well with our data:
The $V$ magnitudes show a mean difference of $0.03 \mbox{~mag}$, the $B-V$
colour of the common objects of less than $0.01 \mbox{~mag}$. This not only
provides a good external control of the quality of our photometry, but their
results of stars A, B, C, D, E, I--6, and III--5, which were overexposed on
all our images, could replenish our photometry. To discriminate between
these and our measurements these stars are marked with open triangles in the
CMD in Fig.~\ref{n4755cmd} (and as well in Fig.~\ref{n4755cmdhaufen}) which
contains all stars for which both $B$ and $V$ magnitudes could be
determined. With Table \ref{n4755cdsphot} we provide a list of the entire
photometric data -- excluding the Dachs \& Kaiser (\cite{dachs}) stars.
Again, we added a reference system for stellar numbers -- this time taken
from Arp \& van Sant (\cite{arp4755}). These numbers are encrypted in the
following way: Numbers $<100$ stand for the stars A to Q, the first digit of
the numbers $>100$ represents the quadrant. The two final digits are the
running numbers in the notation of Arp \& van Sant (\cite{arp4755}). This
follows the nomenclature of the WEBDA database (Mermilliod \cite{webda}).

\refstepcounter{table} \label{n4755cdsphot}

Similar to NGC\,4103, the CMD of NGC\,4755 contains a number of stars in a
structure parallel to the lower part of the main sequence. We will discuss
these objects in Sect.~\ref{n4755sect}.

\section{Astrometry} \label{suedastro}

We used the observations from the Carte du Ciel (CdC) project (Brosche
\cite{brosche}) as first epoch plates for the proper motion studies. The CdC
was initiated in 1887 with the goals to provide a catalogue of the entire sky
down to $11^{\rm th}$ magnitude (which later on became the ``Astrographic
Catalogue'') and a sky chart reaching as far as $14^{\rm th}$ magnitude. To
achieve this goal, the sky was divided into declination zones which were
assigned to different observatories worldwide. The observations were made with
astrographs on photographic plates of a size of $16 \mbox{~cm} \times 16
\mbox{~cm}$ covering a field of view of approx.\ $2^\circ \times
2^\circ$. The CdC archives provide a -- for certain regions of the sky --
unique source for proper motion studies, since it is easy to achieve
differences between first and second epoch material of 100 years and more.

The regions of our star clusters belong to the declination zone of Sydney
Observatory. The fields of the Sydney zone are labelled according to
the number of the central star in an internal numbering system. All three
clusters are located somewhat near the edge of the fields. As a consequence,
NGC\,4755 can be found on two different fields. Details about the fields that
cover the objects are summed up in Table \ref{cdcfields}. Different systems of
running numbers, which can be distinguished by super- and subscripts added,
were used for the individual plates. In contrast to other CdC zones, most of
the Sydney plates were not exposed in the typical way with three pointings
forming the characteristical triangle pattern for each star (see, e.g.,
Ortiz-Gil et~al.\ \cite{ortizcdc}): Among the plates used for our studies we
only found one plate with this property ($567_{{\rm RH}}$). Some of the
others, too, show up to three observations, but with different exposure times
and with the stellar images collocated in one line.

\begin{table}
\caption[]{\label{cdcfields} Carte du Ciel fields covering the open star
  clusters NGC\,4103, NGC\,5281, and NGC\,4755. The field numbers were defined
  by the staff of Sydney Observatory}
\begin{tabular}{lrcccrrr}
\hline
cluster & field & \multicolumn{3}{c}{$\alpha$ (2000)} &
\multicolumn {3}{c}{$\delta$ (2000)} \\
 & &  h &  m &  s & $^\circ$ & $\arcmin$ & $\arcsec$ \\
\hline
NGC\,4103 & $1096$ & $12$ & $13$ & $03$ & $-61$ & $31$ & $50$ \\
NGC\,4755 & $ 904$ & $12$ & $47$ & $36$ & $-59$ & $30$ & $57$ \\
          & $ 905$ & $12$ & $59$ & $47$ & $-59$ & $30$ & $43$ \\
NGC\,5281 & $1282$ & $13$ & $51$ & $17$ & $-63$ & $27$ & $45$ \\
\hline
\end{tabular}
\end{table}

To replenish the sparse second epoch material, two additional plates of each
field were observed with Perth Observatory's astrograph. Table
\ref{suedplatten} gives an overview of all the plates used for this work.

Some of the first epoch plates from the Sydney archive were measured manually
with Perth Observatory's Carl Zeiss plate measuring machine, the others were
digitised with APM at Cambridge, UK, the second epoch plates from Perth with
Tautenburg Plate Scanner (TPS, see Brunzendorf \& Meusinger \cite{TPS,
  TPS99}), all with a linear resolution of $10 \mbox{~} \mu \mbox{m}$ to $15 \mbox{~} \mu
\mbox{m}$. More details can be found in Table \ref{suedplatten}.

\begin{table}
\caption[]{\label{suedplatten} List of the photographic plates used for the
  proper motion studies. Plate and field numbers are given as well as the date
  of observation and exposure times, if available. We also add whether the
  corresponding plate was measured manually with Perth Obervatory's Carl Zeiss
  measuring machine (PER) or digitised in Tautenburg (TPS) or Cambridge (APM),
  where an asterisk stands for $15 \mbox{~} \mu \mbox{m}$ linear resolution in
  contrast to $10 \mbox{~} \mu \mbox{m}$ for all other plates. The plates of
  the year 1998 were exposed at Perth Observatory, the others were taken from
  the archive of Sydney Observatory}
\begin{tabular}{rrrl}
\hline
No. & date of obs. & $t_{\rm exp}$ & measured\\
    &              & \multicolumn{1}{c}{[min]} & \\
\hline
\multicolumn{4}{c}{Field 1096 (NGC\,4103)}\\
\hline
$2875_{\rm s}$ & 18.3.1896 & $30$ & PER \\
$3360_{\rm s}$ &  3.5.1897 & $30$ & APM \\
$748_{\rm RH}$ &  7.5.1901 & $80$ & APM \\
$7735Sa$       & 27.4.1978 &   ?  & PER \\
$7736Sa$       &  1.5.1978 &   ?  & PER \\
$28600$        & 14.7.1998 & $60$ & TPS \\
$28601$        & 15.7.1998 & $60$ & TPS \\
\hline
\multicolumn{4}{c}{Field 1282 (NGC\,5281)}\\
\hline
$1335_{\rm s}$ & 25.3.1897 & $30$ & APM* \\
$3391_{\rm s}$ &  6.5.1897 & $30$ & APM* \\
$725_{\rm RH}$ & 14.5.1901 & $80$ & APM \\
$567_{\rm RH}^{\rm c}$ & 14.6.1909 & $3 \times 30$ & APM* \\
$28594$        & 20.6.1998 & $60$ & TPS \\
$28597$        & 23.6.1998 & $60$ & TPS \\
\hline
\multicolumn{4}{c}{Field\,904 (NGC\,4755)}\\
\hline
$2893_{\rm s}$ & 14.4.1896 & $30$ & APM* \\
$3324_{\rm s}$ & 29.4.1897 & $60$ & APM* \\
$758_{\rm RH}$ & 18.6.1901 & $80$ & APM \\
$554_{\rm RH}^{\rm c}$ & 18.5.1909 & $60$ & APM (3 exp.)\\
$7792Sa$       & 18.3.1980 & $20$ & APM* \\
$7793Sa$       & 18.3.1980 & $12$ & APM* \\
$28592$        & 15.6.1998 & $60$ & TPS \\
$28595$        & 23.6.1998 & $60$ & TPS \\
\hline
\multicolumn{4}{c}{Field\,905 (NGC\,4755)}\\
\hline
$1412_{\rm s}$ & 16.4.1894 &  ?   & APM* \\
$2860_{\rm s}$ & 13.3.1896 & $45$ & APM \\
$3333_{\rm s}$ & 30.4.1897 & $30$ & APM \\
$954_{\rm RH}$ & 18.4.1902 & $80$ & APM \\
$28593$        & 20.6.1998 & $60$ & TPS \\
$28596$        & 23.6.1998 & $60$ & TPS \\
\hline
\end{tabular}
\end{table}

To further improve the second epoch data, we included the short CCD exposures
in our analysis. The stellar $(x,y)$ positions were extracted from the
CCD frames with DAOPHOT II routines (Stetson \cite{daophot}).

Starting with an input catalogue to provide a transformation for the positions
and proper motions, $(\alpha,\delta)$ coordinates and the proper motions were
computed using an astrometric software package developed by Geffert
et~al.\ (\cite{geffert97}).

To determine the membership probabilities, we selected areas around the
centres of the clusters. The membership probabilities were computed on the
base of the proper motions using the method of Sanders (\cite{sanders}). The
positions of the stars or their magnitudes did not play any role in the
derivation of the membership probabilities. For further details we refer to
Sanner et~al. (\cite{n0581paper}).

During an attempt to derive the proper motions with the help of the Hipparcos
catalogue (ESA \cite{hipp}) we found a strong dependence of the proper motions
from the position of the objects on the plates which could not be corrected
with the comparably small number of Hipparcos stars. Moreover, for NGC\,5281
and NGC\,4755 we found a dependence of $\mu_\delta$ and -- to a much lower
extent -- $\mu_\alpha \cos \delta$ on the stellar magnitudes. This effect was
variable over the field of view, so that we decided to obtain the proper
motions for these two clusters not for the entire plates, but only in the
regions of the clusters themselves where we can assume the variations of this
magnitude equation to be marginal. As our software to determine proper motions
requires an input catalogue, this catalogue had both to be sufficiently dense
to provide enough reference stars in the small field and had to cover the
entire magnitude range of the photographic plates not to have to
extrapolate the magnitude dependence. These conditions -- together with good
astrometric properties -- are only fulfilled by the Guide Star Catalog
(GSC). We used its version 1.2 developed by R\"oser et~al.\ (\cite{gsc}).

For NGC\,4103, a reduction of the data with the ACT (Urban
et~al.\ \cite{act}) or Tycho--2 (H{\o}g et~al.\ \cite{tycho2}) catalogues did
not significantly improve the results derived with the Hipparcos
catalogue. Only when using the GSC 1.2 as the basis of our reduction, the
effect described above could be eliminated well. Only in the very corners of
the plates systematic deviations were still detected, however, these regions
are not of importance for the study of NGC\,4103.

For NGC\,4103 we obtained good results using quadratic polynomials in $x$ and
$y$ for transforming $(x,y)$ to $(\alpha,\delta)$ for the photographic plates
and cubic polynomials for the CCD images, respectively. The analysis of our
data revealed that the magnitude trend in the fields of NGC\,5281 and
NGC\,4755 could be well compensated with the addition of a linear magnitude
term to the polynomial transformation. Adding terms with products of the $x$
or $y$ coordinates with the magnitudes, which are not uncommon for work with
CdC plates (see, e.g., Ortiz-Gil et~al.\ \cite{ortizcdc}),  or terms of higher
order did not further improve the results for any of the clusters. 

GSC 1.2 only contains positional data on the catalogue stars, but no proper
motions. Therefore our results are only relative proper motions.

\begin{table}
\caption[]{\label{suedpmcentres} Centres of the proper motion distribution of
  cluster and field stars of the three clusters. Note that these values are
  the relative proper motions as determined from the proper motion
  studies. For the computation of absolute proper motions, see
  Sect.~\ref{suedastro}. The ``$\pm$'' values correspond not to errors of the
  measurements, but to the standard deviation of the Gaussian distributions
  fitted to the data}
\begin{tabular}{lcc}
\hline
object    & $\mu_\alpha \cos \delta$ & $\mu_\delta$ \\
          & $[\mbox{mas~yr}^{-1}]$   & $[\mbox{mas~yr}^{-1}]$ \\
\hline
NGC\,4103 & $+0.91 \pm 1.4$ & $+0.36 \pm 1.4$ \\
field     & $-0.92 \pm 5.0$ & $+0.27 \pm 5.0$ \\
\hline
NGC\,5281 & $-0.70 \pm 2.1$ & $+0.67 \pm 2.1$ \\
field     & $-4.36 \pm 7.0$ & $-1.08 \pm 7.0$ \\
\hline
NGC\,4755 & $+0.18 \pm 1.7$ & $-0.32 \pm 1.7$ \\
field     & $-1.71 \pm 6.5$ & $-0.99 \pm 6.5$ \\
\hline
\end{tabular}
\end{table}

\subsection{NGC\,4103} \label{n4103astro}

With the method described above, we determined the positions and
proper motions of a total of 4,379 stars in the entire $2^\circ \times 2^\circ$
field of the photographic plates. The standard deviations of the computed
positions from the GSC 1.2 were of the order of $\sigma=0 \farcs 25$ in both
$\alpha$ and $\delta$ indicating a good transformation of the plate
coordinates. The internal dispersion of the proper motions was computed from
the deviations of the positions from a linear fit of $\alpha$ and $\delta$ as
functions of time. We derived mean values of $\sigma=1.4 \mbox{~mas~yr}^{-1}$
for both $\mu_\alpha \cos \delta$ and $\mu_\delta$ for individual stars.
A list of all proper motions derived from the plates containing NGC\,4103
is presented in Table \ref{n4103cdseb}.

\refstepcounter{table} \label{n4103cdseb}

\begin{figure*}
\centerline{
\includegraphics[width=\textwidth]{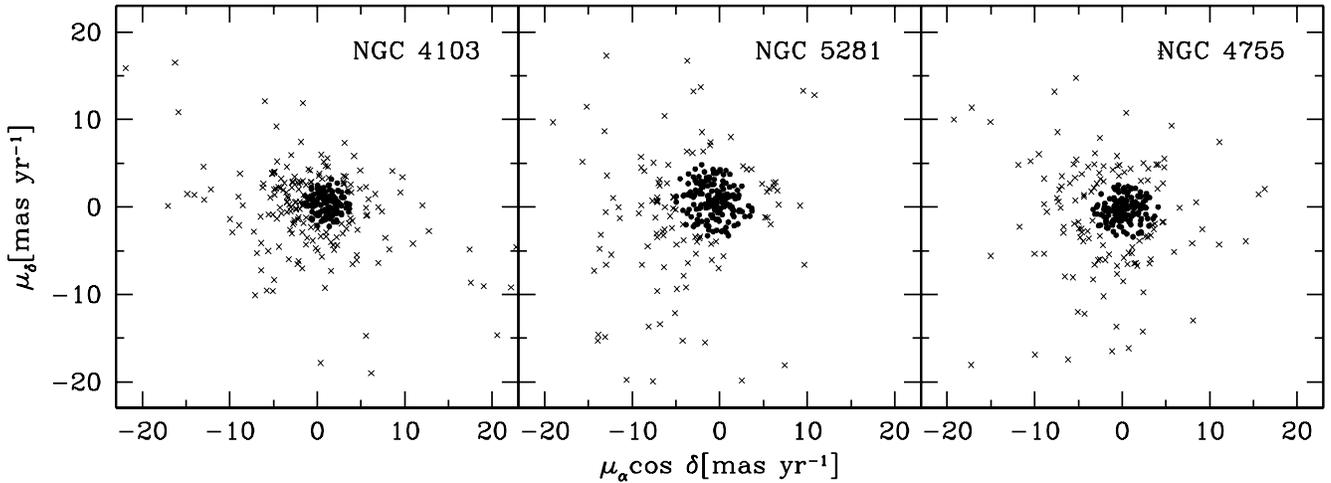}
}
\caption[]{\label{suedvppd} Vector point plot diagrams of the relative proper
  motions found for the stars in the regions of the three clusters. The stars
  which are considered as members according to their proper motions are marked
  with dots, the others with crosses. The stars of NGC\,5281 are not as highly
  concentrated as the objects of the two other clusters. This reflects the
  larger errors of the measurements mentioned in Sect.~\ref{n5281astro}. Note
  the coincidence of the centres of the field and cluster proper motions for
  all three clusters}
\end{figure*}

We selected a $24 \arcmin$ wide field containing 311 stars around
the centre of the cluster to determine the membership probabilities. This
field is large enough to cover all members of the star cluster down to
the limiting magnitude of the plates of around $V=15$ mag. The resulting
vector point plot diagram is shown in Fig.~\ref{suedvppd} (left diagram).

With the standard deviation of the proper motions of $1.4 \mbox{~mas~yr}^{-1}$ as the width of the Gaussian distribution of the cluster stars, we determined the membership probabilities. As the separation
point between members and non-members we chose a probability of $0.6$, leading
 to 128 members and 183 non-members. The values we derived for the
 centres of the proper motion distributions for NGC\,4103 and its surrounding
 field are given in Table \ref{suedpmcentres}.

As the field and cluster proper motions are very similar, we might have
mis-identified several cluster stars as field objects and vice versa. This
cannot be excluded without using further information about the stars
(e.g.\ radial velocities), but it can be expected that the further
investigations are only marginally affected.

To determine the absolute proper motion of NGC\,4103, we compared
our relative proper motions with the Tycho--2 catalogue (H{\o}g et
al.\ \cite{tycho2}). Tycho--2 proved to be a reliable source for stellar proper
motions; its systematic difference to the Hipparcos system is marginal
(Makarov et~al.\ \cite{makarov}, Sanner \& Geffert \cite{tychopaper}). We
determined a shift between the absolute proper motions of Tycho--2 and our
data. Applying this shift to our relative proper motion centre given above, we
computed the proper motion of NGC\,4103 to be
\begin{eqnarray}
\mu_\alpha \cos \delta &=&-6.4 \pm 4.6 \mbox{~mas~yr}^{-1}\\
\mu_\delta &=&+ 0.3 \pm 3.9\mbox{~mas~yr}^{-1}.
\end{eqnarray}
These values correspond well with the results of Baumgardt
et~al.\ (\cite{baudetwie}) who determined the proper motions of open star
clusters from the Hipparcos data.

The sample of our proper motions is complete (except for a few objects which
are too close to a neighbour which made a correct determination of their
positions impossible) down to $V=14$ mag.

\subsection{NGC\,5281} \label{n5281astro}

The area selected around NGC\,5281 has a diameter of $14 \arcmin$, large enough
to cover the entire cluster. The final output catalogue contained 314 stars
showing average uncertainties in proper motions of the order of $2.1 \mbox{~mas~yr}^{-1}$. As for NGC\,4103 the mean difference between the results and
the positional GSC data were approx.\ $0 \farcs 25$ for both coordinates. The
resulting positions and (relative) proper motions are listed in Table
\ref{n5281cdseb}.

\refstepcounter{table} \label{n5281cdseb}

The centres of the proper motion distributions of the cluster and the
surrounding field are again given in Table \ref{suedpmcentres}.
The standard deviation of the Gaussian fit to the distribution of the members
in the vector point plot diagram is $2.1 \mbox{~mas~yr}^{-1}$ in both
coordinates, which equals the $1 \sigma$ error of the proper motions of
individual stars. With Sanders' (\cite{sanders}) method we found a clear
separation between field and cluster stars: Only approx.\ $40$ stars show
membership probabilities of between $0.2$ and $0.8$. We computed the proper
motions for $172$ stars as cluster members by their membership probability of
$0.7$ or higher, and $142$ as field stars. The corresponding vector point plot
diagram can be found in Fig.~\ref{suedvppd}.

Except for four stars (all having close neighbours) the proper motion study
is complete down to $V=14 \mbox{~mag}$, as we found by comparing the results
with the photometric data.

Comparing our relative proper motions with the contents of the Tycho--2
catalogue (H{\o}g et~al.\ \cite{tycho2}), we were able to compute a shift from
our values to the Hipparcos system leading to an absolute proper motion of
NGC\,5281 of
\begin{eqnarray}
\mu_\alpha \cos \delta &=& -7.3 \pm 4.8 \mbox{~mas~yr}^{-1}\\
\mu_\delta &=& -2.0 \pm 4.3\mbox{~mas~yr}^{-1}.
\end{eqnarray}
These values do not coincide with the results of Glushkova et
al.\ (\cite{glush}) who proposed an absolute proper motion for NGC 5281 of
$\mu_\alpha \cos \delta = -17.6 \pm 2\mbox{~mas~yr}^{-1}$ and $\mu_\delta =
-6.4 \pm 1.8 \mbox{~mas~yr}^{-1}$. However, there have been discrepancies of
their results with other studies before (see Sanner et al.\ \cite{capaper}) and
their data were gathered from various sources in the pre-Hipparcos era. We
find better coincidence with the results of Baumgardt et~al.\
(\cite{baudetwie}). There is a difference of approx.\ $2 \mbox{~mas~yr}^{-1}$
in both $\mu_\alpha \cos \delta$ and $\mu_\delta$, however, this discrepancy
is still within the errors of our study.

\subsection{NGC\,4755} \label{n4755astro} 

Advantageously, NGC\,4755 was covered by two neighbouring fields in the Carte
du Ciel survey (see Table \ref{cdcfields}). However, the star cluster is
located almost at a corner of the photographic plates, so that the stellar
images were quite elongated due to distortion effects. We identified
377 stars in the restricted area. These results are listed in Table
\ref{n4755cdseb}. With the plate material available we were able to determine
the proper motions with an accuracy of $\sigma=1.27 \mbox{~mas~yr}^{-1}$ in
both $\mu_\alpha \cos \delta$ and $\mu_\delta$ for individual objects. The
width of the proper motion distribution of the cluster in the vector point
plot diagram (Fig.~\ref{suedvppd}, right diagram) was slightly higher in this
case ($\sigma=1.7 \mbox{~mas~yr}^{-1}$, see also Table \ref{suedpmcentres}),
however, with star counts of the field and cluster areas we were able to
confirm the number of members found.

\refstepcounter{table} \label{n4755cdseb}

The determination of a shift from our relative proper motions to the absolute
system of the Hipparcos catalogue turned out to be critical. Even the Tycho--2
catalogue which is claimed to be complete down to $V=11 \mbox{~mag}$ (H{\o}g
et~al.\ \cite{tycho2}) only contains six stars in common with our proper motion
study. From these objects we derived a shift of $\Delta \mu_\alpha \cos
\delta=-3.06 \pm 3.7 \mbox{~mas~yr}^{-1}$, $\Delta \mu_\delta=-1.01 \pm 4.1
\mbox{~mas~yr}^{-1}$ leading to an absolute proper motion of NGC\,4755 of
\begin{eqnarray}
\mu_\alpha \cos \delta &=& -2.9 \pm 3.9\mbox{~mas~yr}^{-1},\\
\mu_\delta &=& -1.3 \pm 4.3 \mbox{~mas~yr}^{-1}.
\end{eqnarray}
$\mu_\delta$ is almost the same value as found by Baumgardt
et~al.\ (\cite{baudetwie}), for $\mu_\alpha \cos \delta$ there is a difference
of $2 \mbox{~mas~yr}^{-1}$ between their and our results, but within the
errors this still shows a good conformity.

For three Hipparcos stars in the field of NGC\,4755 we were unable to
determine proper motions due to crowding on the photographic plates. With the
derived shift from our relative proper motions to the Hipparcos system we
found that all three objects are likely cluster members. Therefore they are
included in the CMD of the cluster members (Fig.~\ref{n4755cmdhaufen}).

\section{Analysis of the colour magnitude diagrams} \label{suedcmds}

To analyse the CMDs of the three clusters the diagrams have to be corrected
for the presence of field stars. For the part of the CMD for which the proper
motion studies are complete, we applied the membership probabilities found in
Sect.~\ref{suedastro} to discriminate between field and cluster stars. Below
this point, the stellar numbers are high enough to statistically subtract the
field stars. This method is described in Sanner et~al.\ (\cite{n0581paper}).

It has to be remarked that for all three clusters the completeness is higher
in the outer fields than in the central cluster regions which contain a high
number of bright stars (see Figs.~\ref{n4103bild}, \ref{n5281bild},
\ref{n4755bild}, and \ref{suedcompl}, \ref{n4755compl},
respectively). We solved this problem, which is only of marginal importance
for the isochrone fitting process, but essential for the determination of the
luminosity and mass functions, as follows: For the field star
subtraction, we ``recalibrated'' the number of stars in the outer fields to the
completeness of the cluster region in each block of the CMD by
\begin{equation}
N'_{\rm B}=N_{\rm B} \cdot \frac{C_{\rm A}}{C_{\rm B}},
\end{equation}
when $N_{\rm B}$ is the measured star number in the outer field in an
arbitrary block, $N'_{\rm B}$ the corrected star number in the same field, and
$C_{\rm A}$ and $C_{\rm B}$ the completeness factors in the inner and outer
fields, respectively. This correction means that the outer field completeness
is artificially reduced to the values derived for the cluster region. In addition, we had to apply weights to compensate the different sizes of the
areas under consideration. Thus, we can subtract the field from the cluster
region in a 1:1 ratio, while later on, for the determination of the luminosity
and mass functions, the remaining data can be treated for incompleteness in
the usual way -- but with the completeness of the inner cluster area. This
means that we disimproved the completeness of the outer regions, but the
limits for the LF and the IMF computation are set by the data with least
completeness, anyway. This method was preferred to a full correction to $100
\%$ completeness already at this point, since we would have had to deal with
additionally added ``fake'' stars in the CMD.

With the figures corrected in this way, we statistically subtracted the field
stars, obtaining a CMD with the completeness function of the cluster region.

After the statistical field star subtraction stars that immensely differ from
the cluster main sequence remained in the CMDs. This can be explained with the
low number of stars in these CMD regions and hence imperfect statistics. Stars
which can be excluded to be cluster members had to be manually eliminated
before the determination of the luminosity and initial mass functions. As a
rule, we state that all stars which differ from the path of the isochrone by
more than twice the combined photometric error in magnitude and colour do not
belong to the cluster. For the ``upper right'', i.e.\ the brighter (or redder),
limit we allowed another $0.75$ mag in magnitude to account for probable
double stars which can be lifted by up to this value in the CMD, assuming a
1:1 mass ratio of the two components (see, e.g., Aparicio
et~al.\ \cite{aparicio}). These limits are marked in the corresponding CMDs
with dotted lines.

We fitted isochrones based on the models of Bono et~al.\ (\cite{isoteramo}) and
provided by Cassisi (private communication) to the cleaned $(B,V)$ and -- if
available -- $(V,I)$ CMDs. We mainly used the $(B,V)$ data; with the $(V,I)$
CMDs we only checked the quality of the isochrone fits by comparing the
parameters found for each diagram.

We computed the luminosity and initial mass functions on the base of the $B$
and $V$ data. The initial stellar masses were computed on the base of the
information provided with the isochrones as $6^{\rm th}$ order polynomials
of the $V$ magnitudes:
\begin{equation}
m[M_\odot]=\sum_{i=0}^6 a_i \cdot (V[\mbox{mag}])^i. \label{suedmlreq}
\end{equation}
Polynomials of lower than $6^{\rm th}$ order caused higher deviations
especially in the low mass range. The parameters $a_i$ as determined for the
three clusters are listed in Table \ref{suedmlr}.

\begin{table}
\caption[]{\label{suedmlr}Parameters for the polynomial approximation of the
  mass-luminosity relation for the stars of the three clusters. See
  Eq.\ (\ref{suedmlreq}) for the definition of the parameters $(a_0, \ldots,
  a_6)$}
\begin{tabular}{rr@{$\cdot$}lr@{$\cdot$}lr@{$\cdot$}l}
\hline
 & \multicolumn{2}{c}{NGC\,4103} & \multicolumn{2}{c}{NGC\,5281} & \multicolumn{2}{c}{NGC\,4755}\\
\hline
$a_0$ & $-6.6073$ & $10^{+2}$ & $-5.9106$ & $10^{+2}$ & $-3.4752$ & $10^{+2}$ \\
$a_1$ & $+2.6229$ & $10^{+2}$ & $+2.4727$ & $10^{+2}$ & $+1.5219$ & $10^{+2}$ \\
$a_2$ & $-4.2915$ & $10^{+1}$ & $-4.1196$ & $10^{+1}$ & $-2.6223$ & $10^{+1}$ \\
$a_3$ & $+3.6121$ & $10^{0}$  & $+3.5544$ & $10^{0}$  & $+2.2594$ & $10^{0}$ \\
$a_4$ & $-1.6660$ & $10^{-1}$ & $-1.6876$ & $10^{-1}$ & $-1.0504$ & $10^{-1}$ \\
$a_5$ & $+4.0150$ & $10^{-3}$ & $+4.1980$ & $10^{-3}$ & $+2.5271$ & $10^{-3}$ \\
$a_6$ & $-3.9639$ & $10^{-5}$ & $-4.2866$ & $10^{-5}$ & $-2.4752$ & $10^{-5}$ \\
\hline
\end{tabular}
\end{table}

The IMF was computed with a maximum likelihood
technique. We prefer this method instead of the ``traditional'' fit to a
histogram, since bin sizes and location of the bins may influence the results,
and different weights are put on stars in different bins, as all {\it bins}
obtain equal weights. For a more detailed discussion of this point, see Sanner
et~al.\ (\cite{capaper}).

\subsection{NGC\,4103} \label{n4103sect}

To obtain a CMD of NGC\,4103 from which the basic parameters of the cluster
can be derived, we subtracted the field stars in the way described above: For
stars brighter than $V=14 \mbox{~mag}$ we selected all stars with a
membership probability of $P \ge 0.6$, for the fainter stars we statistically
subtracted the stars of fields B and C from those of field A. Star counts
showed a slight gradient in the field star population across the three fields
observed, so that it was mandatory to use a weighted combination of both
fields for the field star subtraction. On the other hand, it could be proved
as well that in fields B and C the number of stars does not decrease with
increasing distance from the centre of the cluster, whereas the stellar
density is higher in field A. We therefore can assume that field A covers all
cluster members, so that the subtraction of the entire fields B and C is
justified.

\begin{figure*}
\centerline{
\includegraphics[width=\textwidth]{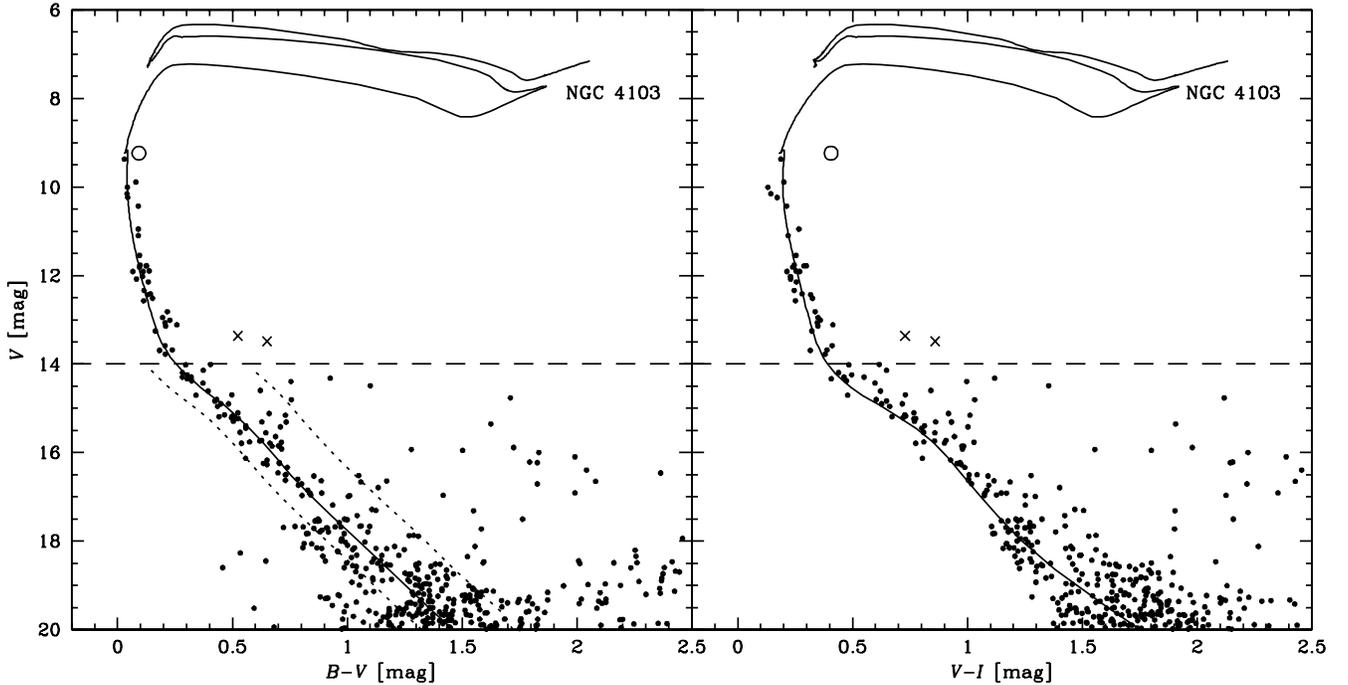}
}
\caption[]{\label{n4103cmdshaufen} Field star corrected colour
  magnitude diagrams of NGC\,4103. The dashed lines indicate the border
  between statistical field star subtraction and membership determination by
  proper motions. The dotted lines in the $(B,V)$ CMD sketch the corridor used
  for the determination of the IMF of NGC\,4103, thus eliminating stars which
  obviously do not belong to the main sequence population of the cluster. The
  stars marked with crosses and a circle are discussed in
  Sect.~\ref{n4103sect}. For the parameters of the fitted isochrone see Table
  \ref{suedparams}}
\end{figure*}

The cleaned CMDs are much more sparsely populated than the original
ones. Especially the feature which seemed to suggest a secondary main sequence
has disappeared. The CMDs with the ``best'' isochrones fitted are shown as
Fig.~\ref{n4103cmdshaufen}. The parameters found
for NGC\,4103 as derived from the isochrone fit are listed in Table
\ref{suedparams}. We found identical values for the $(B,V)$ and $(V,I)$ CMDs,
indicating the good quality of the data and the calibration. Compared with the
findings of Sagar \& Cannon (\cite{sagar4103}), we find good agreement of
these parameters. Only the reddening differs by $\Delta (B-V) \approx 0.05
\mbox{~mag}$.

Two stars in the range of the proper motion study (around $V \approx 13.4$
mag with $B-V \approx 0.6$ mag, marked with crosses in
Fig.~\ref{n4103cmdshaufen}) differ from the main sequence by a few tenths of a
magnitude in colour. These objects were not taken into account for the IMF
investigations. The most likely explanation for their high membership
probability is that these objects are field stars, but by chance show the
cluster proper motion. In addition, located near the completeness limit of the
photographic plates, the inaccuracies of their proper motions might be higher
than average.

After eliminating the non-main-sequence stars as described in
Sect.~\ref{suedcmds}, we determined the LF of NGC\,4103 as a function of the
$V$ magnitude. We present this LF in the left diagram of
Fig.~\ref{suedlkf}. The IMF bases on 170 stars in a mass range from $9.7
M_\odot$ (corresponding to $V=9.24 \mbox{~mag}$) to $0.9 M_\odot$ ($V=18.52
\mbox{~mag}$). The IMF slope $\Gamma$ was computed with our maximum
likelihood software resulting in $\Gamma=-1.46 \pm 0.22$, a result which is in
good agreement with the margin proposed by Scalo (\cite{scalo2}), however,
the IMF appears to be a bit shallower than the average for the mass interval
under consideration. The IMF is plotted in Fig.~\ref{suedimf} (left diagram)
together with a histogram which we added to the plot for reasons of
illustration and as an additional check of the quality of our results.

\begin{figure}
\centerline{
\includegraphics[width=\hsize]{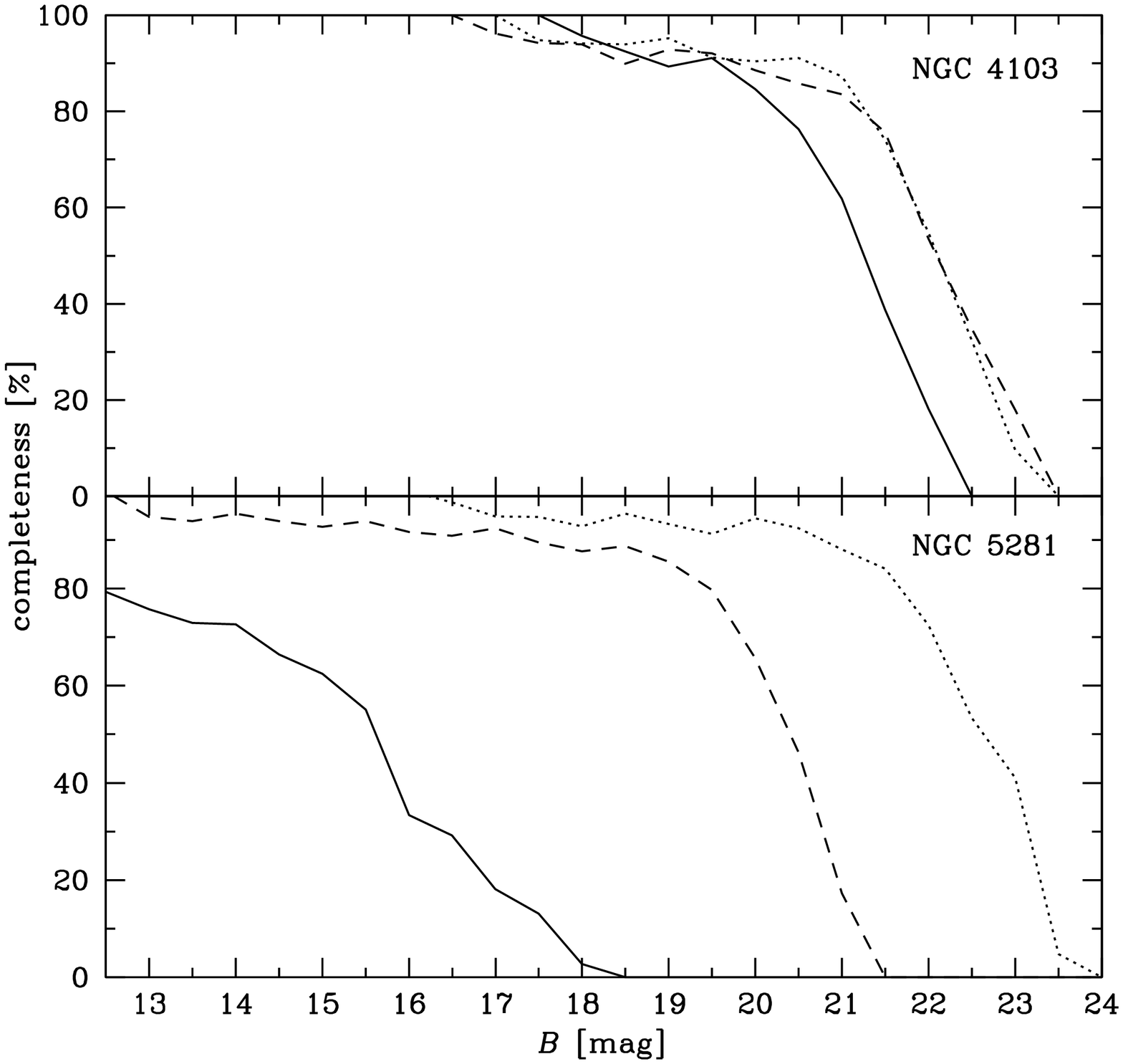}
}
\caption[]{\label{suedcompl} Completeness of NGC\,4103 and NGC\,5281. For
  NGC\,4103 (upper diagram), we present the results for the $900 \mbox{~s}$
  (field A) and $1200 \mbox{~s}$ (fields B and C) $B$ exposures. The solid,
  dotted, and dashed lines correspond to fields A, B, and C, repectively. Note
  that the completeness drops at brighter magnitudes for field A, not only
  because of the shorter exposure time, but also due to the overexposed area
  caused by the brightest stars. This effect is by far more dramatic for
  NGC\,5281 in the lower diagram, where the solid line represents the inner
  $500 \times 500$ pixels of the averaged $600 \mbox{~s}$ $B$ exposures of
  field A and the dotted one the $1200 \mbox{~s}$ images of field B. To get a
  more realistic overview about the completeness of the NGC\,5281 data, we
  also studied the completeness of the five combined $180 \mbox{~s}$ $B$
  exposures of field A (dashed line)}
\end{figure}

Experiments with the lower mass limit of the IMF showed that if the value is
chosen to be smaller than $0.9 M_\odot$, the IMF slope increases with
decreasing mass limit down to $\Gamma=-2.1$ at $m=0.73 M_\odot$. At this
point, the completeness limit of field A is 0.5 (see the upper diagram of
Fig.~\ref{suedcompl}), and 292 stars are covered by this IMF. Since the
maximum likelihood method puts equal weights on every star, the IMF is
dominated by the high number of stars in the region below $1 M_\odot$, so that
the uniform increase of the IMF at higher masses becomes less important.

The effect might have its origin at the different completeness limits of the
photometry. The incompleteness at higher magnitudes is caused by the brightest
stars which cover significant portions of the cluster. In this sense, the
completeness derived before provides only average values for field
A. Assuming mass segregation (i.e.\ a concentration of low mass stars in the
outer regions of the cluster, as proposed, e.g., by Raboud \& Mermilliod
\cite{raboud2}), many low mass stars would be located in the parts where the
completeness is not influenced by the bright stars, so that the incompleteness
in the inner region of NGC\,4103 would be over-compensated by our completeness
correction, resulting in a too steep IMF. Another explanation would be that
NGC\,4103 indeed possesses a higher than average fraction of stars with masses
below $1 M_\odot$ which is in contrast to the shape of the IMF in the $\la 1
M_\odot$ mass range as assumed by Scalo (\cite{scalo2}). This discussion could
only be decided with additional observations.

Stetson (\cite{stetson4103be}) suggests Wesselink's (\cite{wesselink}) star 12
(our star 2; marked with circles in Fig.~\ref{n4103cmdshaufen}) to be a Be
star. The membership probability of $0.66$ gives no clear evidence
about whether this object indeed is a member of NGC\,4103 or not. Its
location in the CMDs supports Stetson's (\cite{stetson4103be}) proposition.

\subsection{NGC\,5281} \label{n5281sect}

The determination of the completeness of the data provided a major challenge
in this study: As a consequence of the bright stars in the centre of the
cluster, large portions of the cluster area were severely overexposed in the
long exposures. Therefore, the completeness of our data sample varies
within field A and from field A to field B, as can be seen in
Fig.~\ref{suedcompl}, lower diagram. With this background we decided to
perform the statistical field star subtraction in two steps: We subtracted
field B, which we corrected for the differing completeness and field size, from
the inner region of field A, the portion which is heavily overexposed, and
from the outer parts of field A separately. Before combining the two field
star corrected samples, we removed further stars of the outer part to simulate
the lower completeness of the cluster region.

\begin{figure*}
\centerline{
\includegraphics[width=\textwidth]{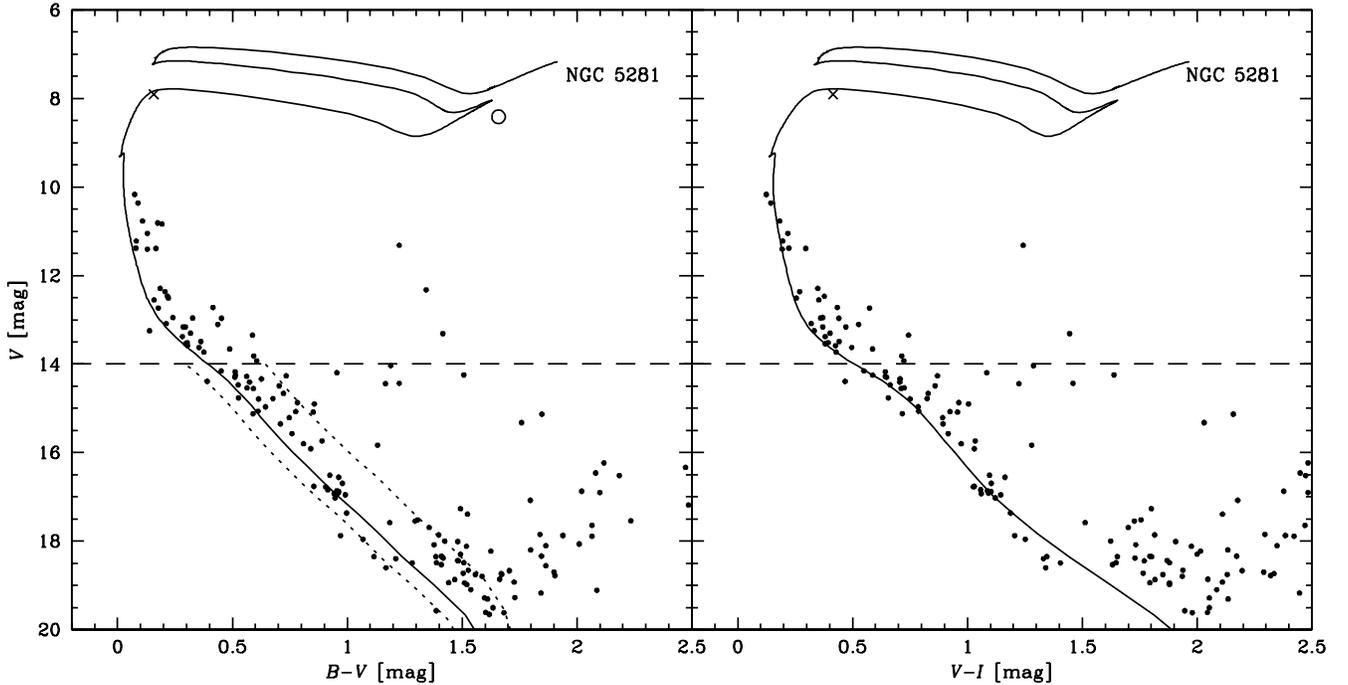}
}
\caption[]{\label{n5281cmdshaufen}Field star corrected colour
  magnitude diagrams of NGC\,5281. The different lines are explained with
  Fig.~\ref{n4103cmdshaufen}. The star marked with a cross was claimed to be a
  blue straggler. The object indeed is a member, however, we do not see
  evidence for a blue straggler nature. The star marked with a circle is
  mentioned in Clari\'{a} (\cite{claria}). For this star, no $I$ brightness
  could be derived due to saturation. See Sect.~\ref{n5281sect} for details}
\end{figure*}

With star counts we found that outside field A the stellar density remains at
a constant level, so that we propose that the entire cluster fits into field
A, at least within the reach of our photometry. This result means that
NGC\,5281 is slightly smaller than proposed in the Lyng{\aa} (\cite{lynga})
catalogue (see the remarks in Sect.~\ref{n5281photometry}). After combining the
members derived from the proper motion study for stars with $V \le 14 \mbox{~mag}$ with the results from the statistical subtraction, we derived the CMDs
of NGC\,5281 presented in Fig.~\ref{n5281cmdshaufen}. The diagrams show that
the cluster is a very sparsely populated object.

In the $(B,V)$ CMD the two brightest stars measured (i.e.\ except for star 1
from the Moffat \& Vogt (\cite{moffatvogt}) sample) have significantly evolved
from the main sequence. These two stars together with the objects with $10
\mbox{~mag} \le V \le 11 \mbox{~mag}$ played an important role in the age
detemination process by our isochrone fit. We found that the $45$ Myr
isochrone best represents both CMDs of NGC\,5281. This isochrone is
overplotted in Fig.~\ref{n5281cmdshaufen}.

The two stars mentioned were subjects of earlier studies: Star 1 (Moffat \&
Vogt \cite{moffatvogt}: star 3) appears in Mermilliod's (\cite{mermiobs}) list
of blue stragglers. Citing this work, Ahumada \& Lapasset (\cite{bluestrag})
included it in their list of blue stragglers, too. The star, marked in
Fig.~\ref{n5281cmdshaufen} with a cross, does not show abnormal behaviour in
the CMDs, however, with a membership probability of $P=0.78$, we have evidence
for this object really to be a member of NGC\,5281. For the other one, star 2
(star 11 in Moffat \& Vogt \cite{moffatvogt}) we can confirm its membership
with $P=0.92$. This star was discussed in Clari\'{a} et~al.\ (\cite{claria})
who declared the object to be a member only from its photometric properties
(see also Clari\'{a} \& Lapasset \cite{claria83}). They discussed the
metallicity of red stars in open clusters and obtained $[{\rm Fe}/{\rm
  H}]=-0.18$ for NGC\,5281, justifying the choice of the Solar metallicity
isochrone sample for the fitting procedure.

The magnitude range $15 \mbox{~mag} \la V \la 18 \mbox{~mag}$ (or $1.3
M_\odot \ga m \ga 0.9 M_\odot$) seems to be very sparsely populated. An
explanation might be that the statistical field star subtraction removed too
many stars in this interval. This effect is also very obvious in the LF, the
middle diagram of Fig.~\ref{suedlkf}. This phenomenon can most easily be
explained by the differing completeness limits in combination with a probable
inhomogeneity of the distribution of the cluster stars: Although we
distinguished between the inner and outer regions of field A, we can assume
the completeness to vary within the central part of the cluster as
well. Assuming most cluster members of this magnitude range to be located in
the part with worst completeness, their number would be
underestimated. Furthermore, a probable mass segregation might decrease this
effect for fainter stars. Under these circumstances, the determination of the
IMF of NGC\,5281 cannot produce reliable results for masses below $m=1.3
M_\odot$. An IMF from $m=1.3 M_\odot$ to the turn-off point of the main
sequence around $V=10.1 \mbox{~mag}$ (corresponding to $m=5.6 M_\odot$) leads
to the slope of $\Gamma=-1.60$ which is about the average value which can be
expected. For this reduced mass interval only 55 members are taken into
account for the IMF, so that the formal error of the maximum likelihood
analysis increases up to $0.5$. We present this result plotted with a
histogram of the mass distribution in Fig.~\ref{suedimf}.

\subsection{NGC\,4755} \label{n4755sect}

\begin{figure}
\centerline{
\includegraphics[width=\hsize]{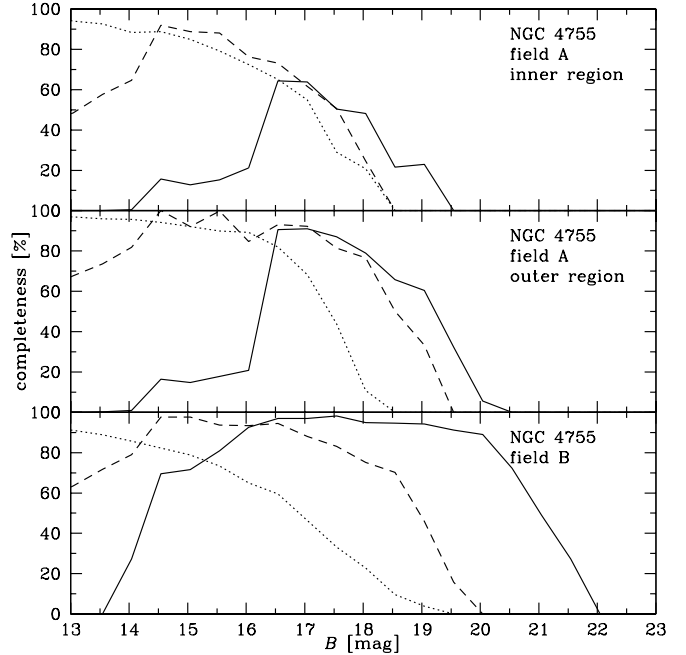}
}
\caption[]{\label{n4755compl} Completeness of the $B$ exposures of
  NGC\,4755 and the surrounding field. The completeness is given for the
 inner region of field A, its outer region, and for field B. The solid line
 stands for the $1800 \mbox{~s}$, dashed lines for the $180 \mbox{~s}$, and
 dotted lines for the $10 \mbox{~s}$ exposures. Note the low sensitivity of
 the inner part of field A due to the bright, overexposed stars in this region}
\end{figure}

Among the clusters under investigation, completeness correction was most critical for NGC\,4755. Therefore we decided to study the completeness for
(composite) images of three different exposure times, and we further
distinguished between the central region of $900 \times 900$ pix.\ around the
centre of the cluster containing the brightest stars, the remaining parts of
the field A exposures, and the part of field B which will be used as the field
star region for the statistical field star subtraction. The results are
portrayed in Fig.~\ref{n4755compl}. As expected, the completeness drops
earliest in the inner region of field A. For the completeness calculations we
used the envelope of the overlapping completeness functions of the short,
medium, and long exposures of the three regions. With the help of the
information gathered for Fig.~\ref{n4755compl} we applied the statistical
field star subtraction in the same way as for NGC\,5281
(Sect.~\ref{n5281sect}).

\begin{figure*}
\centerline{
\includegraphics[width=\textwidth]{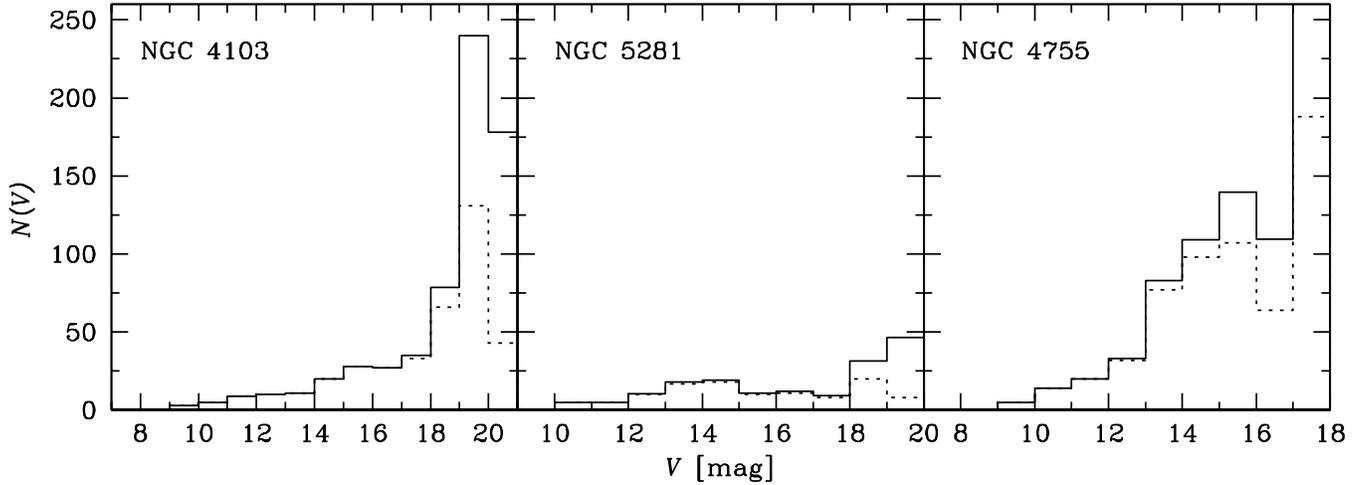}
}
\caption[]{\label{suedlkf} Luminosity functions of the three clusters. Note
  that NGC\,5281 is by far less populated than the two other clusters. For the
  same cluster, a dip in the magnitude range $15 \mbox{~mag} \la V \la 18
  \mbox{~mag}$ is visible, which is discussed in the text
  (Sect.~\ref{n5281sect}). The original data as derived from the photometry
  are indicated with the dotted lines, the solid lines describe the
  completeness corrected LF}
\end{figure*}

Also for NGC\,4755 the proper motion results were complete down to $V \approx
14 \mbox{~mag}$ within field A. Nevertheless, from the combination of cluster
membership according to the proper motions and the statistical subtraction it
appeared that the proper motions for the stars fainter than $V=13 \mbox{~mag}$ were no longer of sufficient quality: Obviously, a too high number of
these stars was excluded from membership, as we found by studying the
LF: At the border between the two methods of field star subtraction the number
of stars climbed from around 20 stars per magnitude up to approx.\ 150 which
appears to be very unlikely. Reducing the membership decision by proper
motions to stars brighter than $13 \mbox{~mag}$ led to a more uniformly
increasing LF (see Fig.~\ref{suedlkf}, right diagram). A source for this
phenomenon might be the higher positional errors for the fainter stars which
could have led to a proper motion distribution which is wider than for the
average. This would mean that a certain amount of faint members is located
outside the cluster section of the vector point plot diagram.

\begin{figure}
\centerline{
\includegraphics[width=\hsize]{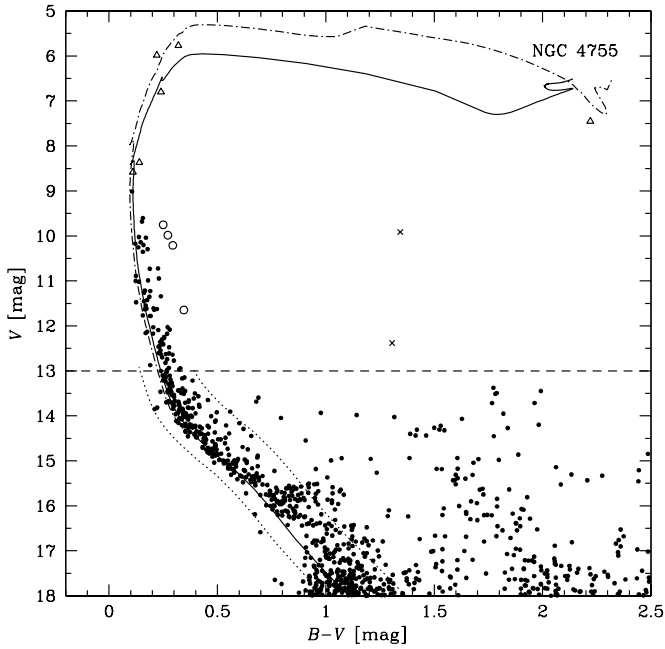}
}
\caption[]{\label{n4755cmdhaufen} Colour magnitude diagram of the members of
  NGC\,4755 with the best fitting isochrone (see Table \ref{suedparams})
  overplotted (solid line). The dash-dotted line in the upper part of the CMD
  is the Geneva isochrone with the same parameters. The other lines are
  explained with Fig.~\ref{n4103cmdshaufen}. The stars marked are discussed in
  Sect.~\ref{n4755sect}}
\end{figure}

In the CMD of the members of NGC\,4755 (Fig.~\ref{n4755cmdhaufen}) as
determined from the proper motion membership analysis and the statistical
subtraction, we found that two red stars had remained in the brighter part
(i.e.\ the part with membership determination by proper motions) of the
CMD. We marked these objects with crosses. Analysing their proper motions we
found that they are located in the vector point plot diagram very close to the
border between members and non-members which provides evidence that their
membership is not very well confirmed. We therefore conclude that these two
objects do not belong to NGC\,4755 and omit them in the following. Four
further objects, which we marked with circles in Fig.~\ref{n4755cmdhaufen},
are located $0.1 \mbox{~mag}$ to $0.25 \mbox{~mag}$ redwards of the main
sequence. For these objects, their nature is less clear. In terms of $V$
magnitudes, the objects are too high {\it above} the main sequence to
represent a binary main sequence. Therefore it seems possible that some of
these stars are Be stars (see, e.g., Zorec \& Briot \cite{zorec}). Indeed,
Slettebak (\cite{slettebak}) lists the three objects around $V=10 \mbox{~mag}$ as Be stars.

In addition, various red stars ($B-V \ga 1.5 \mbox{~mag}$) remained in the
cluster CMD. There would be no physical explanation for these stars if they were
members of the cluster, so that we assume them to have remained due to the
imperfect statistics of the sample. All stars outside a corridor around the
``ideal'' main sequence according to the isochrone fit were not taken into
account for the IMF determination. This corridor is plotted in
Fig.~\ref{n4755cmdhaufen}.

In addition to our photometry, we added those stars from Dachs \& Kaiser
(\cite{dachs}) for which we derived sufficiently high membership probabilities
in the CMD (Fig.~\ref{n4755cmdhaufen}). These stars helped in the isochrone
fitting process. We marked them in the diagram with triangles.

The isochrone with an age of $10 \mbox{~Myr}$
fitted the main sequence best, however, the very red star is slightly too red
for the isochrone created from the Bono et~al.\ (\cite{isoteramo}) model. On
the other hand, older isochrones would no longer represent the brightest blue
stars well. An isochrone from the model of the Geneva group (Schaller
et~al.\ \cite{schaller}) seems to represent the upper part of the CMD
better. The parameters derived from this isochrone are the same as
with the Bono et~al.\ (\cite{isoteramo}) one. For a comparison, we overplotted
the Schaller et~al.\ (\cite{schaller}) isochrone in
Fig.~\ref{n4755cmdhaufen}. Table \ref{suedparams} sums up the parameters found
for NGC\,4755. The results are very similar to previous findings for this
cluster, e.g.\ the values given by Sagar \& Cannon (\cite{sagar4755}).

It becomes clear that NGC\,4755 is the most populous cluster among the targets
of this study: Down to $V \approx 16.9 \mbox{~mag}$, the point at which the
completeness of the inner part of the cluster drops below 50 \%, we found 417
members. Completeness correction would lead to a total number of stars of 494
brighter than this magnitude. The LF in Fig.~\ref{suedlkf} illustrates this
richness of the cluster.

\begin{figure*}
\centerline{
\includegraphics[width=\textwidth]{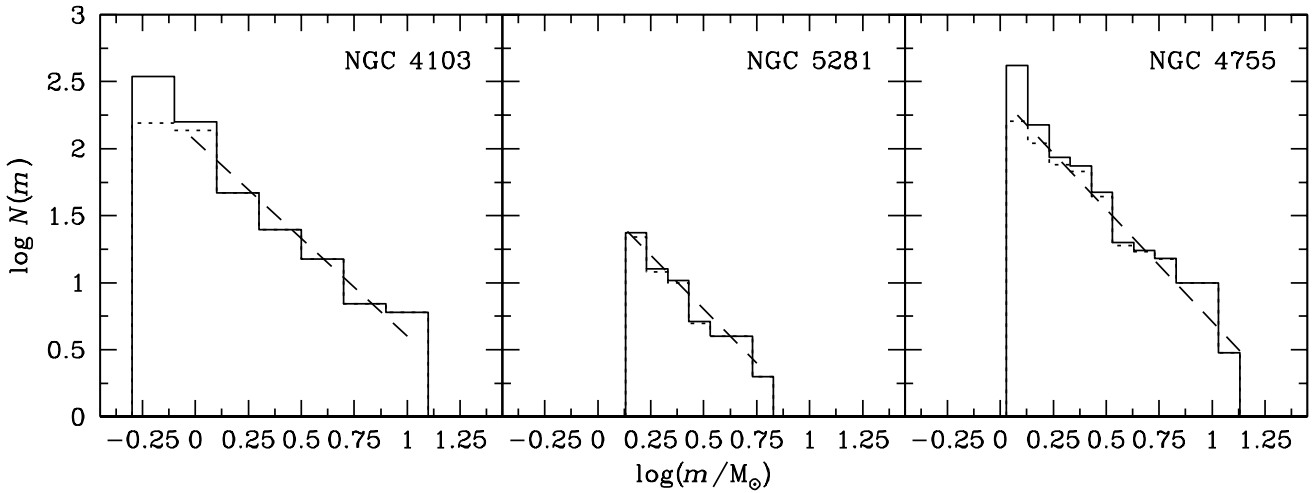}
}
\caption[]{\label{suedimf} Initial mass functions of the three clusters. The
  IMF was computed with a maximum likelihood method; the histogram is only
  plotted for illustration. The line is drawn within the interval for which
  the IMF was determined. The slopes resulting from the completeness corrected
  data are $\Gamma=-1.46 \pm 0.22$ for NGC\,4103, $\Gamma=-1.60 \pm 0.5$ for
  NGC\,5281, and $\Gamma=-1.68 \pm 0.14$ for NGC\,4755}
\end{figure*}

The IMF was computed on the basis of photometric information of these 417
objects, so that it reaches from $m=13.6 M_\odot$ down to $m=1.2 M_\odot$. The
result of $\Gamma=-1.68 \pm 0.14$ represents an IMF slope which almost matches
the average value proposed in Scalo (\cite{scalo2}). We present a plot of the
IMF in Fig.~\ref{suedimf}.

According to Ahumada \& Lapasset (\cite{bluestrag}), six stars of NGC\,4755
are blue stragglers. Three of them (Arp \& van Sant \cite{arp4755} numbers: G,
I--5, and III-1) have membership probabilities that indicate the stars as
non-members, the other three (E, F, and III-5) have sufficiently high
membership probabilities, however, according to their location in the CMD they
do not seem to be blue stragglers. Jakate (\cite{jakate}) found three $\beta$
Cephei stars in the field of NGC\,4755. Two of them, stars F and IV-18, are
members of the cluster, the third one, star I-5, is a field star according to
our proper motion study.

We already mentioned the field stars fainter than $V=14 \mbox{~mag}$ and
approx.\ $1 \mbox{~mag}$ redder than cluster main sequence stars of the same
apparent magnitude (see Sect.~\ref{n4755photo} and Fig.~\ref{n4755cmd}). We
analysed these objects with the help of isochrone fits. It is difficult to
properly fit isochrones to such a mix of objects of different distances and
ages, however, some restrictions could be applied at least to the value of the
reddening: Assuming that the stars are closer than NGC\,4755, the reddening
has to be smaller than the value determined for the cluster, whereas more
distant objects would have to be associated with a higher reddening. The most
likely explanation for these stars is that they are nearby objects of low
luminosity. Their distance can be approximated to be between $80$ and $250
\mbox{~pc}$. Some of the stars might be distant, highly reddened stars with
high absolute magnitude, but their number can be assumed to be small. For a
more detailed investigation of these field stars, additional information
(e.g.\ spectroscopic data) would be necessary.

\section{Concluding remarks} \label{suedsumm}

We studied the initial mass function of the three open star clusters
NGC\,4103, NGC\,5281, and NGC\,4755 based on photometric and astrometric
data. All three IMFs are consistent with a power law shape in the mass range
under investigation, the exponents range from $\Gamma=-1.46$ to
$\Gamma=-1.68$, which is in agreement with typical values as proposed, e.g.,
by Scalo (\cite{scalo2}). The basic parameters of the clusters which were
derived by fitting isochrones to the CMDs are listed in Table
\ref{suedparams}. Together with our previous studies (Sanner
et~al.\ \cite{n0581paper,capaper}) we can summarise that the IMF of all
clusters investigated is universal in the sense that they show the shape of a
power law, however, the slopes of these power laws differ.

\begin{table}
\caption[]{\label{suedparams} Parameters of the three open clusters as derived
  from isochrone fitting to the colour magnitude diagrams}
\begin{tabular}{lr@{ = }l}
\hline
\multicolumn{3}{c}{NGC\,4103}\\
\hline
distance modulus & $(m-M)_0$ & $ 11.7 \pm 0.2 \mbox{~mag}$\\
i.e. distance    & $r$       & $ 2188 \pm 200 \mbox{~pc}$\\
reddening        & $E_{B-V}$ & $ 0.26 \pm 0.02 \mbox{~mag}$\\
age              & $t$       & $ 20 \pm 5 \mbox{~Myr}$\\
i.e.             & $\log t$  & $ 7.3 \pm 0.1$\\
metallicity      & $Z$       & $ 0.02$\\
\hline
\multicolumn{3}{c}{NGC\,5281}\\
\hline
distance modulus & $(m-M)_0$ & $ 11.0 \pm 0.2 \mbox{~mag}$\\
i.e. distance    & $r$       & $ 1580 \pm 150 \mbox{~pc}$\\
reddening        & $E_{B-V}$ & $ 0.20 \pm 0.02 \mbox{~mag}$\\
age              & $t$       & $ 45 \pm 10 \mbox{~Myr}$\\
i.e.             & $\log t$  & $ 7.65 \pm 0.1$\\
metallicity      & $Z$       & $0.02$\\
\hline
\multicolumn{3}{c}{NGC\,4755}\\
\hline
distance modulus & $(m-M)_0$ & $11.6 \pm 0.2 \mbox{~mag}$\\
i.e. distance    & $r$       & $2100  \pm 200\mbox{~pc}$\\
reddening        & $E_{B-V}$ & $0.36  \pm 0.02 \mbox{~mag}$\\
age              & $t$       & $10 \pm 5 \mbox{~Myr}$\\
i.e.             & $\log t$  & $7.0  +0.2/-0.3$\\
metallicity      & $Z$       & $0.02$\\
\hline
\end{tabular}
\end{table}

One aspect of this work was to study the suitability of the Sydney
Carte du Ciel plates as first epoch material for proper motion studies. Our
research showed that in principle the precision of the stellar images on the
plates proved to be satisfactory, although a magnitude dependence of the
positions is present. Therefore it is problematic to derive absolute proper
motions, since a sufficiently dense input catalogue is necessary which covers
the entire magnitude interval of the plates (down to approx.\ $14 \mbox{~mag}$). Such a catalogue is not available yet, it will only be provided by
future satellite missions like DIVA (R\"oser et~al.\ \cite{diva}) and FAME
(Horner et~al.\ \cite{fame}). On the other hand, since the completeness levels
of these catalogues are claimed to be fainter than for the old photographic
plates, the importance of these old archives -- at least for positional
astronomy -- will significantly decrease. They will only be suitable for
astrometric measurements for which long epoch differences are mandatory, like
the search for binary stars which can be revealed by proper motions varying
with time (Wielen et~al.\ \cite{wielen}), or studies in other fields of
astronomy (for examples, see, e.g., Brosch et~al.\ \cite{brosch}).

\begin{acknowledgements}
The authors acknowledge Greg Lowe from Perth Observatory for observing the
second epoch plates for the proper motion study. Further thanks go to Alain
Fresneau, Alan Vaughan, and Elisabeth Moore for initiating the digitisation of the first epoch plates in Cambridge and providing the pre-reduced data
of the plates digitized with the APM. Alan Vaughan is acknowledged for the
access to the Sydney and Melbourne Carte du Ciel archives, Andrea Dieball and
Klaas S.~de Boer for carefully reading the manuscript of this
publication. J.S.\ thanks James Biggs and the entire staff of Perth Observatory
for the permission to use the observatory's plate measuring machine and a very
pleasant stay in Bickley. J.B.\ acknowledges financial support from the
Deutsche Forschungsgemeinschaft under grant ME 1350/3-2. This research has
made use of NASA's Astrophysics Data System Bibliographic Services, the CDS
data archive in Strasbourg, France, and J.-C.\ Mermilliod's WEBDA database of
open star clusters.
\end{acknowledgements}


\begin{thebibliography}{}
\bibitem[1995]{bluestrag}
Ahumada J., \& Lapasset E.\ 1995, A\&AS, 109, 375
\bibitem[1990]{aparicio}
Aparicio A., Bertelli G., Chiosi C., \& Garcia-Pelayo J.\ M.\ 1990, A\&A, 240, 262
\bibitem[1958]{arp4755}
Arp H.\ C., \& van~Sant C.\ T.\ 1958, AJ, 63, 341
\bibitem[2000]{baudetwie}
Baumgardt H., Dettbarn C., \& Wielen R.\ 2000, A\&AS, 146, 251
\bibitem[1959]{becker}
Becker W.\ 1959, Zeitschr. Astroph., 48, 279
\bibitem[1997]{isoteramo}
Bono G., Caputo F., Cassisi S., Castellani V., \& Marconi M.\ 1997, ApJ, 479, 279
\bibitem[2000]{brosch}
Brosch N., Hudec R., Kroll P., \& Tsvetkov M.\ 2000, Saving astronomical
treasures, in A new era of wide field astronomy (in press)
\bibitem[1998]{brosche}
Brosche P.\ 1998, Extracting the information content of the Carte du Ciel
plates, in IV International Workshop on Positional Astronomy and Celestial
Mechanics, ed.\ A.\ L\'{o}pez Garc\'{\i}a, L.\ Yagudin, M.\ Mart\'{\i}nez
Us\'{o}, et~al. (Observatorio Astron\'{o}mico de la Universitat de
Val\'{e}ncia, Val\'{e}ncia), 47
\bibitem[1995]{extinkt}
Burki G., Rufener F., Burnet M., et~al.\ 1995, A\&AS, 112, 383
\bibitem[1998]{TPS}
Brunzendorf J., \& Meusinger H.\ 1998, Astrometric properties of the Tautenburg
Plate Scanner, in The Message of the Angles -- Astrometry from 1798 to 1998,
ed.\ P.\ Brosche, W.\ R.\ Dick, O.\ Schwarz, et~al. (Verlag Harri Deutsch,
Thun, Frankfurt a.M.), 148
\bibitem[1999]{TPS99}
Brunzendorf J., \& Meusinger H.\ 1999, A\&AS, 139,141
\bibitem[1983]{claria83}
Clari\'{a} J.\ J., \& Lapasset E.\ 1983, JA\&A, 4, 117
\bibitem[1989]{claria}
Clari\'{a} J.\ J., Lapasset E., \& Minniti D.\ 1989, A\&AS, 78,363
\bibitem[1984]{dachs}
Dachs J., Kaiser D.\ 1984, A\&AS, 58, 411
\bibitem[1997]{hipp}
ESA 1997, The Hipparcos and Tycho Catalogues, ESA SP--1200
\bibitem[1997]{geffert97}
Geffert M., Klemola A.\ R., Hiesgen M., \& Schmoll J.\ 1997, A\&AS, 124, 157
\bibitem[1997]{glush}
Glushkova E.\ V., Zabolotskikh M.\ V., Rastorguev A.\ S., Uglova I.\ M., \& Fedorova A.\ A.\ 1997, PAZh, 23, 90
\bibitem[2000]{tycho2}
H{\o}g E., Fabricius C., \& Makarov V.\ V.\ 2000, A\&A, 355, L27
\bibitem[1998]{fame}
Horner S.\ D., Germain M.\ E., Greene T.\ P., et~al.\ 1998, BAAS, 193, 1206
\bibitem[1978]{jakate}
Jakate S.\ M.\ 1978, AJ, 83, 1179
\bibitem[1992]{landolt}
Landolt A.\ U.\ 1992, AJ, 104, 340
\bibitem[1987]{lynga}
Lyng{\aa} G.\ 1987, Catalog of open cluster data, $5^{\rm th}$ edition
\bibitem[1979]{king4103}
King D.\ S.\ 1979, Journ. Proc. RSNSW, 112, 13
\bibitem[1980]{king4755}
King D.\ S.\ 1980, Journ. Proc. RSNSW, 113, 65
\bibitem[2000]{makarov}
Makarov V.\ V., Odenkirchen M., \& Urban S.\ 2000, A\&A, 358, 923
\bibitem[1982]{mermiobs}
Mermilliod J.-C.\ 1982, A\&A, 109, 37
\bibitem[1999]{webda}
Mermilliod J.-C.\ 1999, WEBDA: Access to the Open Cluster Database, in
Very Low-Mass Stars and Brown Dwarfs, ed.\ R.\ Rebolo (Cambridge Univ. Press,
Cambridge) (in press)
\bibitem[1973]{moffatvogt}
Moffat A.\ F.\ J., \& Vogt N.\ 1973, A\&AS, 10, 135
\bibitem[1998]{ortizcdc}
Ortiz-Gil A., Hiesgen M., \& Brosche P.\ 1998, A\&AS, 128, 621
\bibitem[1998]{raboud2}
Raboud D., \& Mermilliod J.-C.\ 1998, A\&A, 333, 897
\bibitem[1998a]{diva}
R\"oser S., Bastian U., de~Boer K.\ S., et~al.\ 1998a, Highlights in Astr., 11,
583
\bibitem[1998b]{gsc}
R\"oser S., Morrison J., Bucciarelli B., Lasker B., \& McLean B.\ 1998b,
Contents, Test Results, and Data Availability for GSC 1.2, in Proc. IAU
Symp. 179, New Horizons from Multi-Wavelength Sky Surveys, ed.\ B.\ McLean,
D.\ A.\ Golombek, J.\ J.\ E.\ Hayes, H.\ E.\ Payne (Kluver, Dordrecht), 420
\bibitem[1995]{sagar4755}
Sagar R., \& Cannon R.\ D.\ 1995, A\&AS, 111, 75
\bibitem[1997]{sagar4103}
Sagar R., \& Cannon R.\ D.\ 1997, A\&AS, 122, 9
\bibitem[1955]{salpeter}
Salpeter E.\ E.\ 1955, ApJ, 121, 161
\bibitem[1999]{n0581paper}
Sanner J., Geffert M., Brunzendorf J., \& Schmoll J.\ 1999, A\&A, 349, 448
\bibitem[2000]{capaper}
Sanner J., Altmann M., Brunzendorf J., \& Geffert M.\ 2000, A\&A, 357, 471
\bibitem[2001]{tychopaper}
Sanner J., \& Geffert M.\ 2001, A\&A (submitted)
\bibitem[1971]{sanders}
Sanders W.\ L.\ 1971, A\&A, 14, 226
\bibitem[1986]{scalo1}
Scalo J.\ M.\ 1986, Fund. Cosm. Phys., 11, 1
\bibitem[1998]{scalo2}
Scalo J.\ M.\ 1998, The IMF revisited -- A case for variations, in ASP
Conf. Series 142 , ed.\ G.\ Gilmore, D.\ Howell (ASP, San Francisco), 201
\bibitem[1992]{schaller}
Schaller G., Schaerer D., Meynet G., \& Maeder A.\ 1992, A\&AS, 96, 269
\bibitem[1985]{slettebak}
Slettebak A.\ 1985, ApJS, 59, 769
\bibitem[1981]{stetson4103be}
Stetson P.\ B.\ 1981, AJ, 86, 1500
\bibitem[1991]{daophot}
Stetson P.\ B.\ 1991, Initial experiments with DAOPHOT II and WFC images, in
$3^{\rm rd}$ ESO/ST-ECF Garching Data Analysis Workshop, ed.\ P.\ J.\
Grosb{\o}l, R.\ H.\ Warmels (ESO, Garching), 187
\bibitem[1998]{act}
Urban S.\ E., Corbin T.\ E., \& Wycoff G.\ L.\ 1998, AJ, 115, 2161
\bibitem[1969]{wesselink}
Wesselink A.\ J.\ 1969, MNRAS, 146, 329
\bibitem[1999]{wielen}
Wielen R., Dettbarn C., Jahrei\ss\ H., Lenhardt H., \& Schwan, H.\ 1999, A\&A,
346, 675
\bibitem[1991]{zorec}
Zorec J., \& Briot D.\ 1991, A\&A, 245, 150
\end{thebibliography}
\end{document}